\documentclass[10pt,letterpaper]{article}
\usepackage[top=0.85in,left=2.75in,footskip=0.75in]{geometry}
\usepackage{graphicx} % Required for inserting images
\usepackage{cite}
\usepackage{nameref,hyperref}
\usepackage{xcolor}
\usepackage{url}
\usepackage{doi}

\usepackage{float}
\usepackage[german,english]{babel} % English and German language 
\usepackage{booktabs} % Horizontal rules in tables 
\usepackage{comment} % Necessary to comment several paragraphs at once
\usepackage[utf8]{inputenc} % Required for international characters
\usepackage[T1]{fontenc} % Required for output font encoding for international characters
\usepackage{amsmath,amsfonts,amsthm} % Math packages which might be useful for equations
\usepackage{tikz} % For tikz figures (to draw arrow diagrams, see a guide how to use them)
\usepackage{tikz-cd}
\usetikzlibrary{positioning,arrows} % Adding libraries for arrows
\usetikzlibrary{decorations.pathreplacing} % Adding libraries for decorations and paths
\usepackage{tikzsymbols} % For amazing symbols ;) https://mirror.hmc.edu/ctan/graphics/pgf/contrib/tikzsymbols/tikzsymbols.pdf 
\usepackage{blindtext} % To add some blind text in your paper
\usepackage{enumerate}
\usepackage{xcolor}
\usepackage{algorithm}
\usepackage{algorithmic}
\usepackage{subcaption}

\usepackage[aboveskip=1pt,labelfont=bf,labelsep=period,justification=raggedright,singlelinecheck=off]{caption}

\makeatletter
\renewcommand{\@biblabel}[1]{\quad#1.}
\makeatother

\usepackage{lastpage,fancyhdr,graphicx}
\usepackage{epstopdf}
\fancyheadoffset[L]{2.25in}
\fancyfootoffset[L]{2.25in}
\usepackage[right]{lineno}

\begin{document}
\vspace*{0.2in}

\begin{flushleft}
{\Large
\textbf\newline{
Shaping Opinions in Social Networks with Shadow Banning
}
}
\newline
\\
Yen-Shao Chen\textsuperscript{1*} and
Tauhid Zaman\textsuperscript{1}
\bigskip

\textbf{1} School of Management, Yale University, New Haven, Connecticut, United States of America
\bigskip

% Use the asterisk to denote corresponding authorship and provide email address in note below.
* Corresponding author 

Email: yen-shao.chen@yale.edu
\end{flushleft}

% \begin{document}
% \maketitle
\section*{Abstract}
The proliferation of harmful content and misinformation on social networks necessitates content moderation policies to maintain platform health.  One such policy is \emph{shadow banning}, which limits content visibility. The danger of shadow banning is that it can be misused by social media platforms to manipulate opinions. Here we present an optimization based approach to shadow banning that can shape opinions into a desired distribution and scale to large networks.  Simulations on real network topologies show that our shadow banning policies can shift opinions and increase or decrease opinion polarization. We find that if one shadow bans with the aim of shifting opinions in a certain direction, the resulting shadow banning policy can appear neutral.  This shows the potential for social media platforms to misuse shadow banning without being detected.  Our results demonstrate the power and danger of shadow banning for opinion manipulation in social networks.

%%%%%%%%%%%%%%%%%%%%%%%%%%%%%%%%%%%%%%%%%%%%%%%%%%%%%%%%%%%%%%%%
% \linenumbers
\section{Introduction}
%Social networking platforms host big discusssions
%Platforms also have risks, such as spreading disinformation or exacerabting polarization in a society.  
The digital age has borne witness to the rapid rise of social networks which influence the dynamics of public conversation. Inherent to their structure and expansive reach, these platforms possess the potential to shape public discourse, often wielding influence that transcends geographical boundaries. However, this powerful capacity can also serve as a conduit for the propagation of harmful content or disinformation.  The ramifications of this can be significant, including societies being ensnared in a web of misinformation, or becoming perilously polarized to the brink of internal conflict. This necessitates strategies designed to stymie the potential exploitation and misuse of these influential platforms.

%if its bad, ban, if its kinda bad, shadow ban
For content that manifestly constitutes a threat, platforms have the authority to expunge the user responsible. This course of action is typically employed in scenarios involving explicit threats of violence, unequivocal disinformation posing potential danger, or other instances that breach the platform's stipulated policies. However, not all content resides within such clear-cut boundaries of propriety. 
 
Certain types of content may straddle the periphery of policy violation without crossing the explicit threshold. Even though such content does not transgress the policies directly, its rampant dissemination can subtly skew the tenor of online discourse in ways that could engender undesirable outcomes. Consider, for instance, an ongoing political debate on the platform marked by heightened tension and polarization. In such circumstances, the platform may deem it necessary to curb the spread of emotionally-charged content that could further inflame the situation and potentially instigate violent acts. The content in question here may not represent a clear-cut violation of the platform's policies. Nonetheless, its unchecked proliferation could exacerbate polarization at a critical juncture, thereby posing potential risks.

%shadow banning
To address content that does not directly breach the platform's rules but still presents certain risks, the platform can employ various content moderation strategies. One of these strategies is referred to as \emph{shadow banning}.  Characterized by its clandestine nature, shadow banning operates by limiting the visibility of a user's content, effectively curtailing the user's reach without their awareness  \cite{suzor2019we}. Shadow banning allows platforms to exert control over the content they host, without disrupting user engagement significantly.  It can be employed at different levels of precision.  The platform could limit the visibility of all content of a user, or it could be more selective and limit the visibility of the user's content to a set of specific users.  In either case, the net effect is that certain content posted by a user will not be seen by others.

%shadown banning can piss people off, must be used sparingly
Shadow banning can serve as an effective strategy for maintaining the health of social media discourse. It was found that Twitter shadow banned accounts that exhibited automated or bot-like behavior, along with offensive posts \cite{jaidka2023silenced}. While there are obvious benefits to shadow banning,  it is not without potential drawbacks. Of significant concern is the potential to upset users who may perceive this practice as a form of covert censorship, infringing on their freedom of expression. Given the clandestine nature of shadow banning, users may feel betrayed or manipulated upon discovering that their content reach has been curtailed without their knowledge. These sentiments could lead to decreased user engagement, trust erosion, and potentially a mass exodus from the platform.  For instance, many conservative users of Twitter accused the platform of shadow banning them as an exercise of political censorship \cite{savolainen2022shadow}. Instagram has been accused of disproportionately shadow banning women  in an attempt to limit the spread of inappropriate content, accusations which Instagram denies \cite{cook2019instagram}.  The negative stigma surrounding shadow banning has caused Elon Musk, owner of Twitter, to publicly state that he will not allow shadow-banning on the platform \cite{musto2023musk}. Thus, while the judicious application of shadow banning policies can be effective at content moderation, it is imperative that such measures are deployed sparingly and transparently. This careful balancing act between user satisfaction and content moderation underscores the intricate challenge of managing contemporary social media platforms.

%lit review platform bias
The danger of content moderation policies such as shadow banning is that they can result in the manipulation of opinions by the platform.  Traditionally, opinion manipulation has been considered from the perspective of a user in the network.  The goal is to select target users to receive content in order to maximize an objective, such as the reach of the content \cite{kempe2003maximizing, kempe2005influential,leskovec2007cost,chen2009efficient,chen2010scalable,aral2018social}  or the mean opinion in the network \cite{yildiz2013binary,hunter2022optimizing,ghezelbash2019polarization}.  Other approaches to opinion manipulation focus on increasing the credibility of opinion leaders  \cite{Zhao2018}.   In contrast, content moderation is done by the platform itself and does not introduce new content into the network.  Rather, it modifies the audience for existing content.  However, content moderation can still manipulate opinions.  For instance, one type of content moderation is recommendations, where the platform uses an algorithm to choose what content to show users.  The recommendation algorithm is typically designed to show users content they are likely to prefer.  Many studies have found that the bias of content recommendation algorithms creates a positive feedback loop that can lead to increased polarization \cite{sirbu2019algorithmic, peralta2021effect, perra2019modelling, blex2022positive, iannelli2022filter, cinus2022effect}. 
%add reference to visibility limitation
It has been found that uniformly limiting the visibility of content in a social network can also inadvertently increase polarization \cite{vilela_majority-vote_2021}. From these results it is clear that content moderation can manipulate opinions in a social network, even when this was not the intention.

%When social network platforms employ shadow banning, or any type of  content moderation, there can be unintended consequences. One type of content moderation is  content recommendation, where the platform uses an algorithm to choose what content to show users.  The recommendation algorithm is typically biased to show users content they are likely to prefer. Content recommendation can be viewed as the opposite of shadow banning, as it results in an increase in the content seen in the network. Many studies have found that the bias of content recommendation algorithms creates a positive feedback loop that can lead to increased polarization \cite{sirbu2019algorithmic, peralta2021effect, perra2019modelling, blex2022positive, iannelli2022filter}.  The authors in \cite{cinus2022effect} show that increased polarization can also arise from algorithms that recommend people to follow instead of content to view. The important conclusion from these results is that content moderation can manipulate opinions in a social network, even when this was not the intention. For instance, content recommendation algorithms are designed to show users content they like, but  they unintentionally polarize opinions. 

The ability of content moderation to affect opinions is very concerning.  It raises the question of whether or not a social media platform could design content moderation policies with the explicit objective of manipulating the opinions into an arbitrary target distribution.  If this was possible, it could be very dangerous for a society.  Furthermore, one can ask whether this opinion manipulation could be done without being detected.  For instance, can a social media platform deploy content moderation methods, like shadow banning, with a partisan intent, yet still uphold an outward semblance of political neutrality? This scenario suggests that a society could be covertly swayed by a social media platform, with the populace remaining unaware until potentially harmful consequences have firmly taken hold.

%here we show how to find optimal shadown banning policies
In this work, we demonstrate how a social media platform can employ a different form of content moderation, specifically shadow banning, to arbitrarily shape the opinions of its users.  We frame this as an optimization problem which allows one to calculate  shadow banning policies that shape opinions into a specified target distribution. The shadow banning policy is obtained by solving a simple linear program.  Because of this, our approach can scale to large networks and can accommodate a variety of opinion dynamics models encompassing complex  phenomena such as bounded confidence \cite{hegselmann2002opinion}.  When determining shadow banning policies, we focus on two principal characteristics of the opinion distribution: the mean and the variance.

Altering the mean enables the platform to steer the prevalent sentiment regarding a topic in a designated direction. When utilized with upright intentions, this can permit the platform to curtail the spread of hazardous content and diminish the influence of misinformation. However, if employed with unethical intentions, manipulating the mean may allow the platform to forge an artificial bias either in favor or against a particular topic. This holds a potential for substantial risk, especially if, for example, implemented during an election year. 

In contrast, manipulating the variance does not generate a bias towards a topic, but instead alters the overall character of the online dialogue. Reducing the variance has the potential to moderate online polarization and suppress intense sentiment. On the other hand, amplifying the variance enhances polarization and escalates the severity of sentiment. There are scarce justifiable reasons to amplify variance unless the objective is to destabilize a populace via information warfare. Nonetheless, it is an action that can be effortlessly executed through shadow banning. This shows the potential risks of shadow banning and emphasizes that it must be used with great care.

This paper is organized as follows.  We begin by presenting the underlying opinion dynamics model used in our analysis.  We then show how to calculate shadow banning policies by solving a linear program.  Shadow banning policies are calculated for synthetic networks to provide intuition for their behavior.  We then calculate shadow banning policies on two large-scale Twitter networks for multiple opinion objectives.  We find that substantial manipulation of opinions can be achieved over time, even with limited shadow banning.  Finally, we show that if one shadow bans with a politically biased objective in mind, such as maximizing the opinion mean, the resulting shadow banning policy appears to be politically neutral, or biased in a counter-intuitive way.  

%%%%%%%%%%%%%%%%%%%%%%%%%%%%%%%%%%%%%%%%%%%
\section{Methods}
\subsection{Opinion Dynamics Model}
Shadow banning can be used to control the movement of opinions.  However, we must first have a model for the underlying dynamics of the opinions in order to apply shadow banning.  There are a variety of such models in existence, but they can all be reduced to a set of continuous time differential equations.  We now present this differential equation framework and our choice of opinion dynamics model.

We represent the social  network as a directed graph $G = (V,E)$ where $V$ is the set of vertices, which are the users of the social network platform, and $E$ is the set of edges which represent following relationships.  This model is appropriate with social networks with a follower/following structure, such as Twitter, Instagram, or TikTok. An edge $(i,j)$ pointing from user $i$ to user $j$ means that user $j$ follows user $i$, and subsequently will be shown content posted by $i$.  User $i$ posts content to user $j$ at a rate $\lambda_{ij}$, which is the number of posts per unit time. In practice this rate would only depend on $i$ as it would correspond to his posting rate.  However, it is possible that the rate could vary with $j$, if for instance $j$ does not check the platform often and thus does not see all of $i$'s content.  Therefore, we can consider $\lambda_{ij}$ as an effective posting rate from $i$ to $j$.   Also, we will consider shadow banning policies that limit the rate at which content flows along individual edges in the social network, so separating posting rates by edge simplifies our analysis. 

Each user $i$ has a time dependent latent opinion $\theta_i(t)$ which is a real number. The opinion of content posted by a user at any given time matches their latent opinion.  More general models allow for the content to have a random opinion which equals $\theta_i(t)$ in expectation \cite{hunter2022optimizing}.  However, we will not consider such stochastic generalizations here.

Each time a user $i$ posts in the network, all users update their opinions.  Assume $i$ posts at time $t$ and consider a user $j$.  If $j$ does not follow $i$ then there is no change in $j$'s opinion.  However, if $j$ does follow $i$, then $j$ changes his opinion by an amount given by $f(\theta_i(t)-\theta_j(t))$, where $f$ is the \textit{opinion shift function} and its argument is the difference of the opinions of $i$ and $j$.  This form for the opinion shift function is in accordance with many popular opinion dynamics models \cite{degroot1974reaching,hegselmann2002opinion}.  

To simplify this analysis, we approximate the opinions as continuous functions.  This is a good approximation for large networks.  We first assume that users independently post content according to a Poisson process.  Then the number of posts in the entire network is a merged Poisson process of the individual user posting processes.  We define $\delta$ as the mean time between posts in the network.  First consider the case where posts on each edge are independent.  In this case, $\delta = 1/\sum_{(i,j)\in E}\lambda_{ij}$.  Second, consider the more realistic case where users post independently, but their posts are broadcast to all of their followers simultaneously.  In this case $\delta = 1/\sum_{i\in V}\lambda_{i}$ where $\lambda_i$ is the posting rate of user $i$  (and  $\lambda_{ij}=\lambda_i$).  In either case, we see that as the network grows large, $\delta$ becomes increasingly small. Therefore, for large networks, a continuous time approximation is reasonable. We assume there was a post in the network at time $t+\delta$ and write down the update rule for user $j$'s opinion as 
\begin{align*}
    \theta_j(t+\delta) = \theta_j(t) + X_{ij}f(\theta_i(t)-\theta_j(t)).
\end{align*}
The random variable $X_{ij}(t)$ is one if there is a post on edge $(i,j)$ and zero otherwise.  Given that a post occurred, the mean value of $X_{ij}(t)$ is $\lambda_{ij}\delta$ by properties of merged Poisson processes \cite{pishro2014introduction}.  Taking the expectation over $X_{ij}(t)$ and doing some simple manipulations, the update rule becomes
\begin{align*}
    \frac{\theta_j(t+\delta) - \theta_j(t)}{\delta} = \lambda_{ij}f(\theta_i(t)-\theta_j(t)).
\end{align*}
As the network size increases, $\delta$ will approach zero, and the term on the left can be replaced with a time derivative $d\theta_j/dt$.  This then gives us our continuous time opinion dynamics model
\begin{align}
    \frac{d\theta_i}{dt} & = \sum_{j\in V}\lambda_{ji}f(\theta_j-\theta_i)~~ \forall i\in V.
\end{align}
This differential equation model is a good approximation to the opinion dynamics on large networks. In our application, we are considering a platform shadow banning users in the entire social network, so this approximation is valid.  

The last piece to specify in this model is the opinion shift function $f$. There are several options here. The classic DeGroot model has $f(x)=\omega x$ for some non-negative constant $\omega$  which measures how much a single post can shift one's opinion \cite{degroot1974reaching}.  This term is capturing how reliable one considers the opinions of others.  Hearing an opinion from someone deemed more reliable will cause one to change their opinion more than an opinion from someone unreliable.  DeGroot's model leads to opinion consensus on most networks.  This is one flaw of the model, as many researchers have observed persistent polarization in real social networks \cite{adamic2005political, conover2011political, bakshy2015exposure, garimella2018political,rossetti2023bots}.  Another flaw of the model is the fact that the opinion shift is proportional to the difference between the opinion of the post and one's own opinion.  However, it is unlikely that an opinion vastly different from one's own would be persuasive in modern online social media.  Instead, these opinions may be ignored.

To allow for persistent polarization and limit the persuasive power of posts with opinions with vastly different from their audience, the bounded confidence model was proposed \cite{deffuant2000mixing, hegselmann2002opinion}.  In this model the shift function is given by 
\begin{align}
 f(x) = 
\begin{cases} 
\omega x & \text{if } |x| \leq \epsilon \\
0 & \text{otherwise}
\end{cases}
\label{eq:shift}
\end{align}
where $\epsilon$ is the size of the confidence interval.  The bounded confidence model places a limit on the range of trusted opinions.  Opinions deviating too far (by more than $\epsilon$) from one's own opinion  have no persuasive power.   The bounded confidence model can result in consensus or persistent polarization depending upon the value of the confidence interval, the initial opinions, and the network structure \cite{lorenz2006consensus, blondel2009krause, Zhao2018}.  It is a more complex model that better captures behavior in real social networks. In this work we use this bounded confidence model for the opinion dynamics.  We note that there are more complex variations of this model, include labeling community belongings \cite{peng2023role}, factoring in various sociological phenomena \cite{dong2024opinion}, and taking into account opposing opinions when compromise happens \cite{jiang2024analysis}.  Our framework would allow for the utilization of these model variations for the underlying opinion dynamics.

%%%%%%%%%%%%%%%%%%%%%%%%%%%%%%%%%%%%%%%%%%%%%%%%%%%%%%%%%%%%%%
\subsection{Shadow Banning Control}
We can easily incorporate shadow banning into our opinion dynamics model.  We define the shadow banning strength on an edge $(i,j)$ at time $t$ as $u_{ij}(t)$ which is a real number between zero and one.  Shadow banning reduces the posting rate $\lambda_{ij}$ by a multiplicative factor $1-u_{ij}(t)$.  At one extreme, $u_{ij}(t)=1$ corresponds to total censorship of content from $i$ to $j$.  At the other extreme, $u_{ij}(t)=0$ corresponds to no shadow banning. Under shadow banning, the opinion dynamics model is slightly modified to become
\begin{align}
    \frac{d\theta_i}{dt} & = \sum_{j\in V}\lambda_{ji}(1-u_{ji}) f(\theta_j-\theta_i),~~ \forall i\in V. \label{eq:dynamics}
\end{align}
where we have dropped the time arguments to simplify notation.

%implementation
%A shadow banning policy specifies the fraction of content that is allowed to pass on each edge in the network at each time step.  This can be implemented in a variety of ways.  First, we assume that the policy would remain fixed for a certain amount of time.  In practice, this could be one day, week, or month.  The question then becomes how does the platform select which content to block on each edge in a day so that the total blocked fraction is $u_{ij}(t)$ for each edge $(i,j)\in E$.  The challenge is that the posts are created at random times and the total posts in a day is not known a priori.  Using a deterministic approach to shadow banning requires one to know the total number of posts in day in order to determine how many posts to block.  A better approach is to use a randomized algorithm.  Instead of deterministically selecting which posts to block, the platform can randomly block each post on an edge $(i,j)$ independently with probability $u_{ij}(t)$.  On average, this will result in the correct fraction of posts being blocked.  In addition, randomization has the benefit of not needing to know the daily volume of posts ahead of time  and is also simple to implement. 

%%%%%%%%%%%%%%%%%%%%%%%%%%%%%%%%%%%%%%%%%%%%%%%%%%%%%%%%%%%%
%\section{Shadow Banning Optimization}
To determine the shadow banning policy, the social network platform must have a target distribution for the opinions of its users.  This is described by an objective function, or an instantaneous reward, $r(\theta(t))$ of the opinions, where $\theta(t)$ refers to the opinions of each user in the network at time $t$.  The objective can be any function of the opinions, but here we will consider the important cases where the objective is the opinion mean, variance or negative variance.   The negative values allow the platform to minimize the variance under our objective maximization framework.  

The platform can have different types of goals with respect to the objective.  One possible goal is to maximize the objective at a final or terminal time $T$.  This can be formulated as the following control theory problem:
\begin{align*}
    \max_{u} ~~~&r(\theta(T))\\
    \text{subject to} ~~~&\frac{d\theta_i(t)}{dt}=\sum_{j\in V}\lambda_{ji}(1-u_{ji}(t))f(\theta_j(t)-\theta_i(t)), ~~~ \forall i \in V.
\end{align*}
Solving this problem is non-trivial, but could possibly be done using techniques from control theory \cite{evans2005introduction}.  However, there is an issue with scalability.  The shadow banning control problem has one control variable for each edge in the network and one state variable for each user in the network.  If one is performing shadow banning on an entire social media platform, this can result in hundreds of millions of state variables and billions of control variables.  Standard control theory techniques will not work on such large problems.   To avoid this issue, we use the following approximation.  We can rewrite the final objective as
\begin{align*}
    r(\theta(T)) & = \int_{0}^T\frac{dr}{dt}dt.
\end{align*}
An optimal solution to this problem will choose the shadow ban controls in a manner to maximize this integral.  However, the size of the problem prevents such a solution from being found.  A more scalable approach is to find a greedy solution.  Instead of  maximizing the integral, we maximize the integrand at each time step sequentially.   This means we choose the shadow banning policy to maximize  $dr/dt$ at each time $t$.    It turns out that this objective can be maximized in a manner that scales to large networks. To see why, we can rewrite it as
\begin{align*}
    \frac{dr(\theta(t))}{dt} & = \sum_{i\in V}\frac{\partial r}{\partial \theta_i(t)}\frac{d\theta_i(t)}{dt}\\
    & =  \sum_{i\in V}\frac{\partial r}{\partial \theta_i(t)} \sum_{j\in V}\lambda_{ji}(1-u_{ji}(t))f(\theta_j(t)-\theta_i(t))\\
    & = \sum_{(j,i)\in E}B_{ji}(t)(1-u_{ji}(t))
\end{align*}
where we have defined $B_{ji}(t) =\frac{\partial r}{\partial \theta_i(t)}\lambda_{ji}f(\theta_j(t)-\theta_i(t))$.  Above we have used equation \eqref{eq:dynamics} for $d\theta_i/dt$.  We see from this expression that the shadow banning appears linearly in the reward derivative through the opinion dynamics.
This observation gives us an efficient method to find the shadow banning policy: at time $t$ we maximize the time derivative of the instantaneous reward. Because the derivative is a linear function in the shadow ban controls, this maximization problem can be formulated as a linear program.  This allows us to solve for the policy in very large networks using a variety of well-known methods \cite{bertsimas1997introduction, gondzio2012interior, wright1997primal}.

In addition to maximizing the objective, the platform also has constraints on the shadow banning policy.  If the shadow banning is too strong, the user experience will be affected negatively.  Therefore, the constraints limit the strength of the shadow banning.  This limitation can be done at different levels.  One  can set a limit on the mean  shadow banning strength in the entire network, or one can limit the shadow banning strength on individual edges. We refer to these  limits as $s_{network}$ for the network average and $s_{edge}$  for individual edge.  Combining these constraints with the greedy approximation leads to the following linear program for the shadow banning policy at time $t$:
\begin{align*}
\max_{u(t)}~~~ &\sum_{(j,i)\in E} B_{ji}(t)(1-u_{ji}(t)) \\
\text{subject to} ~~~~&\sum_{(j,i)\in E}u_{ji}(t)   
  \leq s_{network} |E| \\
&~~~0 \leq u_{ji}(t) \leq s_{edge}, ~~\forall (j,i)\in E \\
%&~~~\revision{$\frac{d\theta_i}{dt}=\sum_{j\in V}\lambda_{ji}(1-u_{ji}(t))f(\theta_j-\theta_i), ~~ \forall i \in V,$}
\end{align*}
where the decision vector $u(t)$ denotes the set of $u_{ji}(t)$ for every $(j,i)\in E$. $u_{ji}(t)$ indicates the shadow banning strength on the following/follower edge $(j,i)$. The resulting tweet rate on an edge $(j,i)$ then reduces to $\lambda_{ji}(1-u_{ji}(t))$. The coefficients $B_{ji}(t)$ in the objective are determined by the opinions in the network $\theta(t)$, the network structure (which is contained in the edge set $E$), the posting rates $\lambda_{ji}$, the derivative of the reward with respect to the opinions $ \frac{\partial r}{\partial \theta_i(t)}$, and  the opinion shift function $f$.   Because we assume the bounded confidence model for the dynamics, we use the shift function in equation \ref{eq:shift}. As for shadow banning constraints, the first inequality corresponds to the limit of the mean shadow banning strength in the network, while the second inequality corresponds to the limit of the shadow banning strength on each individual edge.

The solution of the linear program gives the shadow ban policy at time $t$.  Solving this linear program at every time step will give the complete dynamic shadow banning policy.  The policy is dynamic because as time progresses, the user opinions change, leading to a potentially different shadow banning policy.  The platform has the flexibility to decide the frequency of policy recalculations, whether it is a daily update or longer intervals like weekly or monthly. To put this policy into action, we can consider $u_{ji}(t)$ as a probability. Therefore, at time $t$, after computing the policy by solving the shadow banning linear program, one potential approach for the platform to implement shadow banning is by making each post from user $j$ invisible to user $i$ with a probability of $u_{ji}(t)$.

The impact of the particular choice of reward function on the resulting shadow banning policy is expressed through the partial derivative of the reward with respect to the opinions.  We list the partial derivatives for the objectives we consider in Table \ref{table:deriv}.  One nice feature of our approach is that the shadow banning policy can be found by solving a linear program for any objective function.  This allows one to use more novel objective functions beyond those considered here.

\begin{table}
\centering
\renewcommand{\arraystretch}{1.75} 
\caption{Table of the partial derivatives for different objective functions.  We have used $\mu$ to refer to the mean of the opinions.}
\label{table:deriv}
\begin{tabular}{|c|c|c|}
\hline
Objective & Objective function $r$ & $\frac{\partial r}{\partial\theta_i}$ \\[0.2cm] \hline
 Maximize mean &     $\dfrac{1}{|V|}\sum_{i\in V}\theta_i$ & $\dfrac{1}{|V|}$ \\[0.2cm] \hline
Minimize variance & $-\dfrac{1}{|V|-1}\sum_{i\in V}(\theta_i-\mu)^2$ & $-\dfrac{2}{|V|-1}(\theta_i-\mu)$ \\[0.2cm] \hline
Maximize variance & $\dfrac{1}{|V|-1}\sum_{i\in V}(\theta_i-\mu)^2$ & $\dfrac{2}{|V|-1}(\theta_i-\mu)$ \\[0.2cm] \hline
\end{tabular}
\end{table}

We can gain insight to the behavior of the shadow banning policy for different objectives by examining the linear program.  Consider an edge $(j,i)$ corresponding to node $i$ following node $j$.  Because we want to maximize the time derivative of the reward, we will shadow ban edges where the coefficient $B_{ji}(t)$ is negative (recall that $u_{ji}(t) = 1$ corresponds to maximum shadow banning).  We first consider the case where the goal is to maximize the opinion mean.  Using the partial derivative of the mean in Table \ref{table:deriv} and the definition of $B_{ji}(t)$, we find that there will be shadow banning on edge $(j,i)$ when $f(\theta_j(t)-\theta_i(t))$ is negative.  The opinion shift functions we consider have odd symmetry and their sign matches the sign of their argument.  This means that there is shadow banning if $\theta_i(t)>\theta_j(t)$.  In this case node $j$ is pulling down the opinion of node $i$, which decreases the opinion mean. Therefore, the  policy shadow bans the edge.  

To understand the variance shadow banning policies, it is useful to define $\mu(t)$ as the mean of the user opinions at time $t$.  If the goal is to minimize the opinion variance, then $B_{ji}(t)$ is negative when the partial derivative of the reward is negative and the opinion shift is positive, or vice versa.   This corresponds to $\theta_i(t) > \mu(t)$  and $\theta_i(t)<\theta_j(t)$, or $\theta_i(t) <\mu(t)$  and $\theta_i(t)>\theta_j(t)$.  In the first case node $i$'s opinion is above the mean and it is being pulled up by node $j$, which increases the variance.  In the other case, $i$'s opinion is below the mean and it is being pulled down by $j$, which also increases the variance.  Therefore, under either of these conditions this edge gets shadow banned.  A similar analysis for maximizing the variance shows that the edges which are shadow banned correspond to a node being above the mean and being pulled down or a node being below the mean and being pulled up.

Comparing the mean and variance policies, we see that the policy for the mean depends only on the opinion shift on an edge since the goal is to shift the distribution in a particular direction.  The global position of an opinion is not relevant.  However, the policy for the variance is more complex since the goal is to either stretch out or compress the opinion distribution around the global mean.  In this case the policy takes into account the position of the opinions relative to the mean in addition to the shift direction on the edge.

%%%%%%%%%%%%%%%%%%%%%%%%%%%%%%%%%%%%%%%%%%%%%%%%%%%%%%%%%%%%%%%%%%%%%%%
\section{Results}
We test our shadow banning algorithm in a variety of networks with different opinion objectives.  We consider maximizing the mean, minimizing the variance, and maximizing the variance.  We first calculate shadow banning policies on small synthetic networks to illustrate some of the intuition for the policies discussed earlier.  We then calculate shadow banning policies on larger Twitter networks to demonstrate the scalability of the algorithm and show how it performs on real network topologies and opinion distributions.  In our analysis we update the shadow banning policies  daily, as this is a practical implementation scheme for social media platforms.

We use the bounded confidence model for the opinion dynamics.  We must choose the parameters $\epsilon$ and $\omega$ to specify the opinion dynamics model.  Larger values for these parameters correspond to stronger persuasion between nodes (wider confidence interval and larger shift magnitude).  We choose to be conservative and choose small values for both parameters to limit the speed of the natural opinion dynamics and keep persistent polarization that is observed in real social networks.   For $\epsilon$ we choose 0.1 so the confidence interval is fairly narrow (the initial opinions are distributed between zero and one). We set $\omega=0.003$, which indicates that users have much more confidence in their own opinions relative to the opinions of others.  This aligns with several studies of persuasion which have found a single message can cause a very small opinion shift in a controlled environment \cite{pink2021elite, chu2021religious, pink2023effects, bai2023artificial}.  We use a value for $\omega$ less than what is implied in these works as we expect there to be many factors that reduce the persuasive power of social media posts, such as user's not seeing a post at all or scrolling past it without reading it. Note that we use constant values for $\epsilon$ and $\omega$ for all users at all time steps. In reality, users are likely to have heterogeneous and time-varying values for these  parameters \cite{vande2016modelling}.  We do not have a good sense of how these parameters are distributed, so we instead choose to use a constant value for all users.  However, if such information was available, it can easily be incorporated into our simulation framework.

%%%%%%%%%%%%%%%%%%%%%%%%%%%%%%%%%%%%%%%%%%%%%%%%%%%%%%%%%%%%%%%%%%%%%%%
\subsection{Synthetic Networks}
\subsubsection{Path Network} 
\begin{figure}[H]
    \centering
    \includegraphics[width=\textwidth]{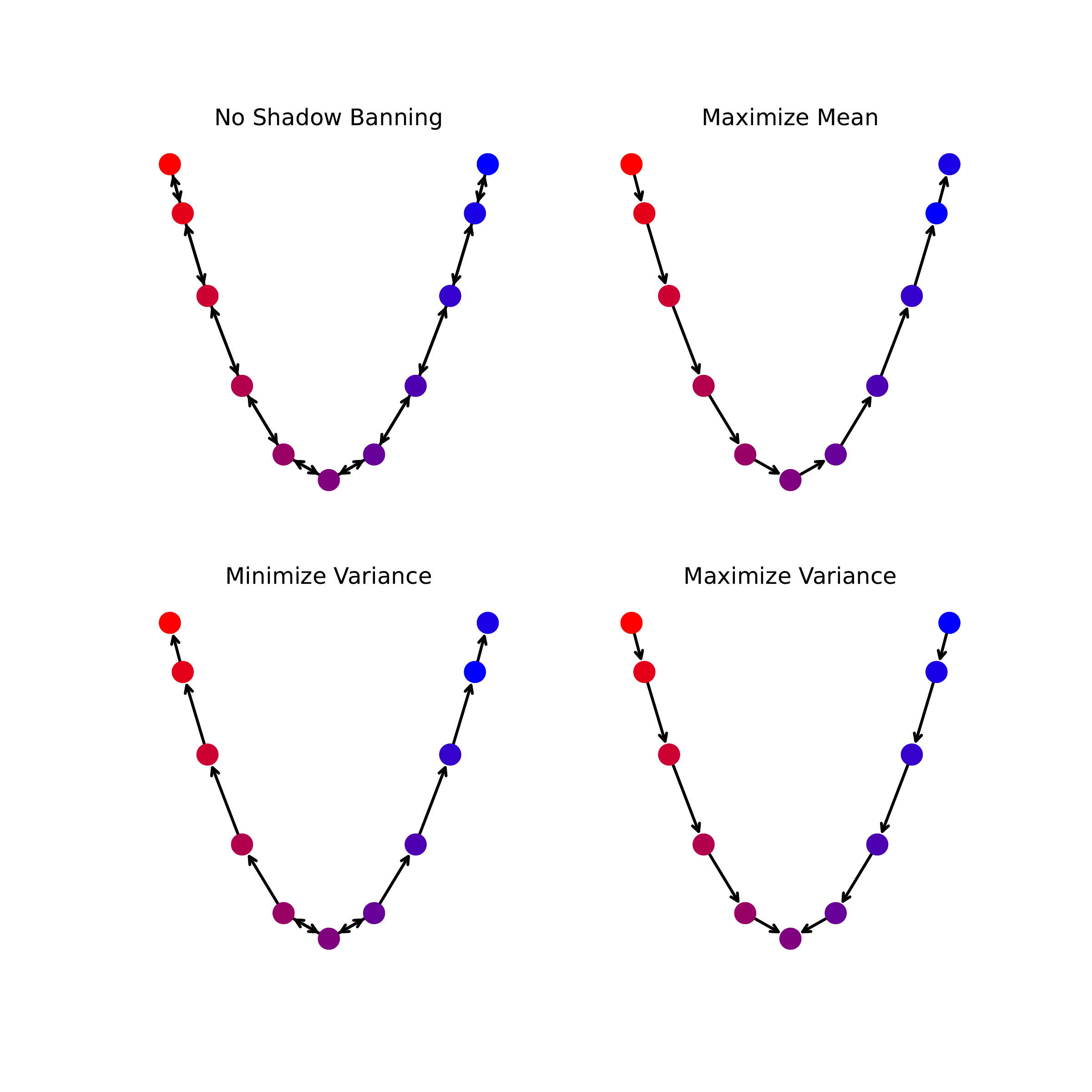}
    \caption{Path network with edges untouched by the initial shadow banning policy for different objectives.  The node colors indicate the opinion (lower are blue, higher are red).  The direction of the edges indicates the flow of information on the network.  The objectives are (top left) no shadow banning, (top right) maximize mean, (bottom left) minimize variance, and (bottom right) maximize variance.  }
    \label{fig:linear_network_policy}
\end{figure}

We begin with a path network shown in Figure \ref{fig:linear_network_policy}.  The network has 11 nodes whose opinions increase linearly from zero at one end to one on the opposite end.  The posting rate of each user is set to one post per day.  Our simulation will run for 365 days, with the shadow banning being updated daily.  We set no limits on the maximum edge shadow banning strength ($s_{edge}=1)$, but we limit the maximum mean shadow banning strength to $s_{network}=0.5$.  To avoid issues with numerical rounding, we set $\epsilon = 0.101$ so the confidence interval strictly greater than the difference between neighboring opinions.

\begin{figure}[H]
    \centering
    \includegraphics[width=\textwidth]{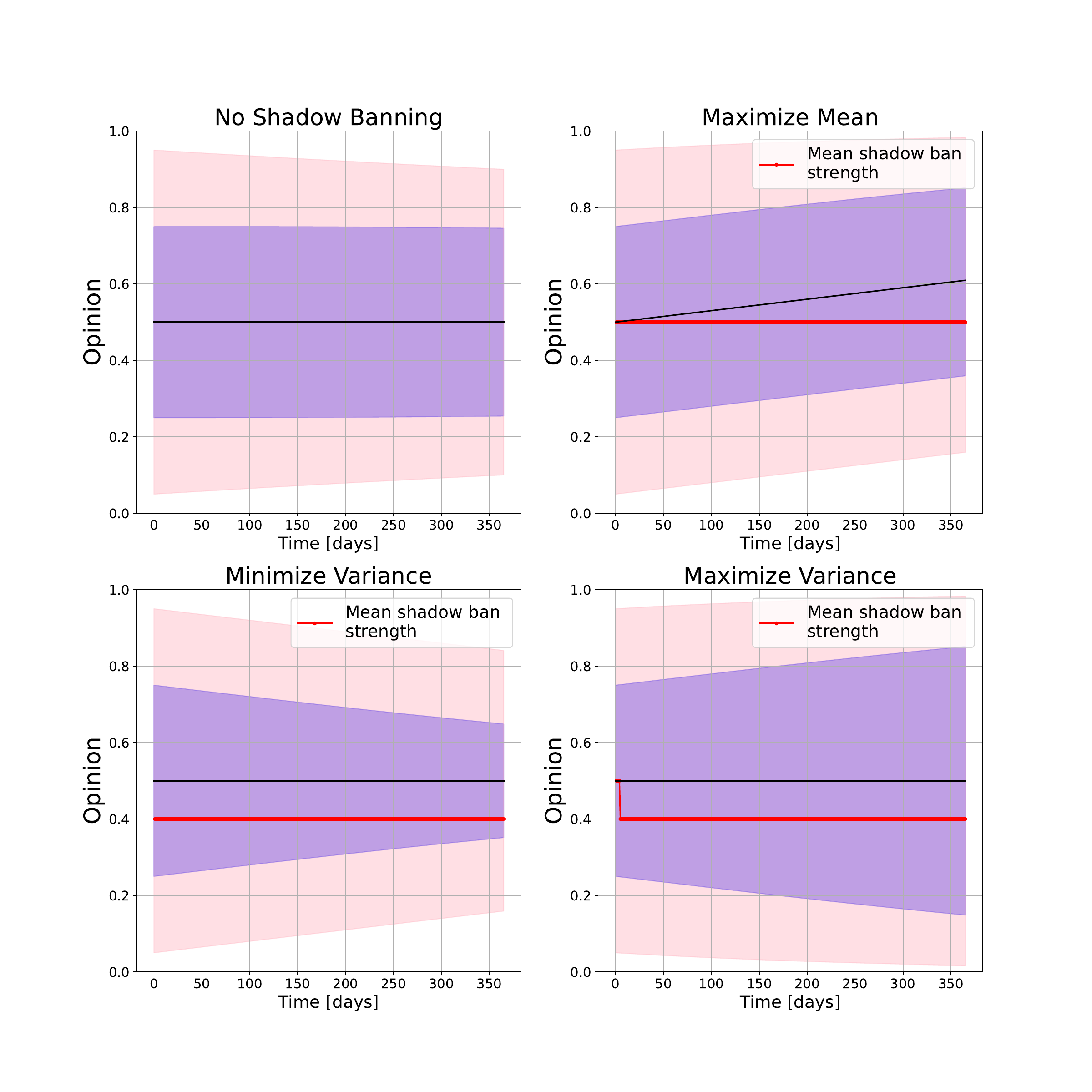}
    \caption{Opinion distributions and mean shadow ban strength versus time under shadow banning control policies for different objective functions on a path network.  For the opinions, the purple region is the 25th to 75th quantiles, and the pink region is the 5th to 95th quantiles.   The objectives are (top left) no shadow banning, (top right) maximize mean, (bottom left) minimize variance, and (bottom right) maximize variance. }
    \label{fig:linear_network_opinions}
\end{figure}

We calculate the shadow banning policy for each objective function and show the evolution of the resulting opinions in Figure \ref{fig:linear_network_opinions}.  With no shadow banning,  the opinions do not converge  and the mean is slightly above 0.5.  When trying to maximize the mean, the opinions are pulled up to  0.6.   For minimizing the variance, we see the opinions converge to 0.5, but with less polarization than with no shadow banning.  For maximizing the variance, the opinions become more polarized as the simulation progresses. For each objective we also show the mean shadow banning strength in the network.  As can be seen, the shadow banning remains near the 0.5 limit set by the linear program throughout the simulations.  This is because the opinions move slowly, so the shadow banning can continue to increase the objective over the simulation duration.

To understand what the different shadow banning control policies are doing for each objective, we visualize the initial decisions ($t=0$) of the shadow banning policy.  We draw the network keeping only non-shadow banned edges  to show where the shadow banning occurred.  We show the resulting networks in Figure \ref{fig:linear_network_policy}.  The structure of these shadow banned networks reflects the intuition from the linear program.  We see that to maximize the mean, the control shadow bans edges pointing from a lower opinion node to a higher opinion node.  This is being done to prevent any nodes from pulling their neighbor opinions down. For minimizing the variance, we see that initially the edges pointing from more extreme opinions are shadow banned, causing the opinions to shift towards the middle.  By having all opinion shifts point towards the middle, the opinions will converge to 0.5 more quickly. For maximizing the variance, the opposite edges are blocked, causing the opinions to drift towards the extremes.

%%%%%%%%%%%%%%%%%%%%%%%%%%%%%%%%%%%%%%%%%%%%%%%%%%%%%%%%%%%%%%%%%
\subsubsection{Stochastic Block Model Network}
Real-world networks exhibit an assortative structure where users of similar opinions exist in distinct clusters in the network, often referred to as echo chambers \cite{adamic2005political, conover2011political, hanna2013partisan, borge2015content, garrett2009echo, garimella2018political, del2016echo, cota2019quantifying, del2016spreading}.  One popular model for this network structure is known as the stochastic block model \cite{holland1983stochastic}.  In this model, one specifies the number of clusters and the number of nodes in each cluster.  Then one specifies  a $k\times k$ probability matrix $p$ where element $p_{ab}$ is the probability of an edge between a node in cluster $a$ and a node in cluster $b$.  All edges are formed independently.  If all values in $p$ are equal, then the stochastic block model reduces to the well-known Erdos-Renyi model \cite{ergraph}.  Generally, the off-diagonal elements of $p$ are less than the diagonal elements to make intra-cluster edges more likely than inter-cluster edges.  This is how the assortativity structure is achieved.

\begin{figure}[H]
    \centering
    \includegraphics[width=\textwidth]{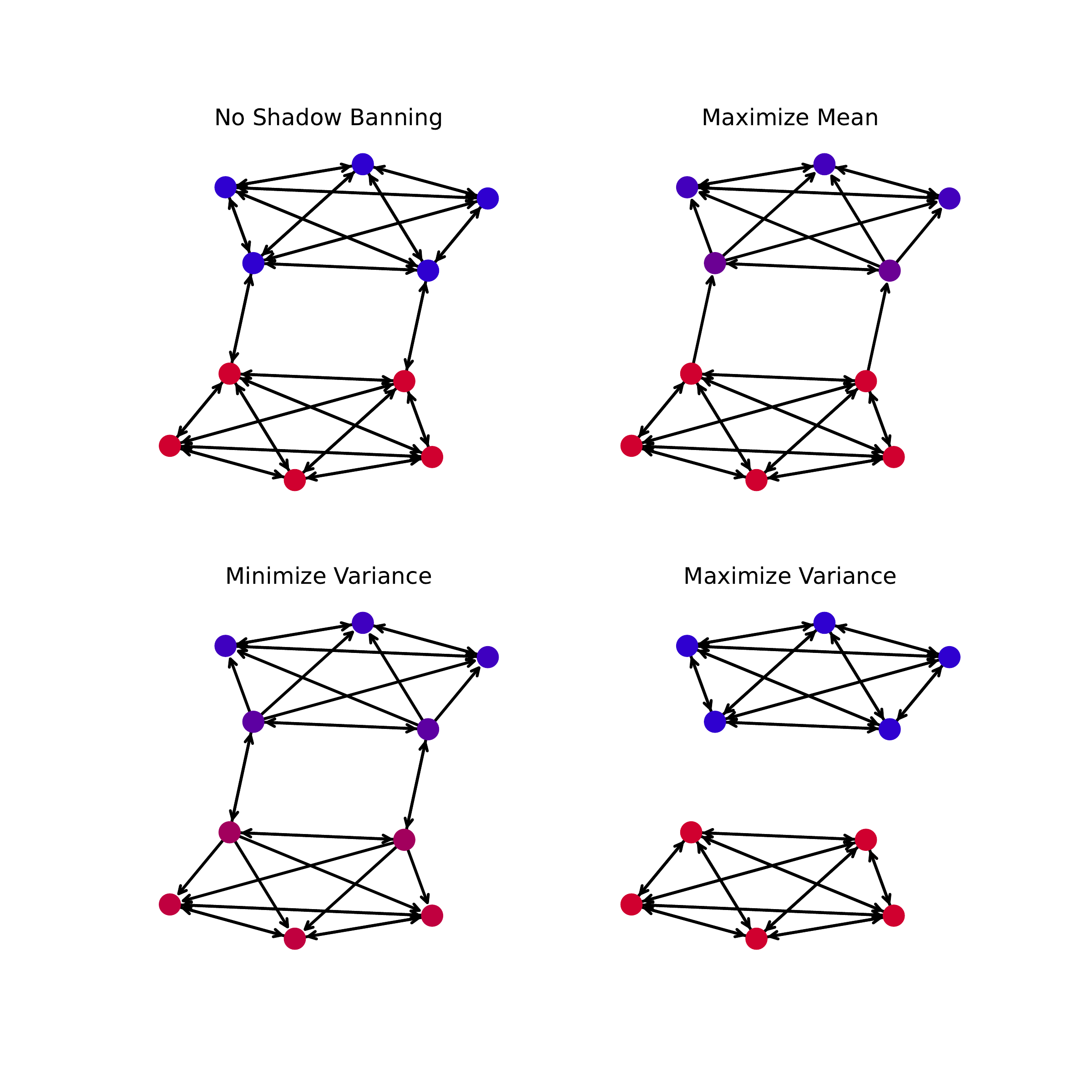}
    \caption{Stochastic block model  network with edges untouched by the  shadow banning policy for different objectives.    The node colors indicate the opinion (lower are blue, higher are red).  The direction of the edges indicates the flow of information on the network.  The objectives are (top left) no shadow banning, (top right) maximize mean, (bottom left) minimize variance, and (bottom right) maximize variance.  For the no shadow banning policy, the node colors correspond to opinions at time  $t=0$.  For the other objectives, the node colors correspond to opinions at time  $t=10$. }
    \label{fig:sbm_network_policy}
\end{figure}

We utilize a stochastic block model network with ten nodes equally divided between two clusters.  The intra-cluster probabilities are one and the inter-cluster probabilities are 0.05.   This produces a network of two cliques connected by a small number of directed edges, as shown in Figure \ref{fig:sbm_network_policy}.  The nodes in each cluster have the same opinion, which is 0.35 in cluster one and 0.65 in cluster two.   We chose these values so that they are close enough to allow some persuasion between the clusters under our model specification.  To allow for non-trivial opinion dynamics, we set $\epsilon$ equal to 0.4.  This allows persuasion to occur between the clusters under natural dynamics.  Otherwise the two clusters do not interact in any meaningful way.

\begin{figure}[H]
    \centering
    \includegraphics[width=\textwidth]{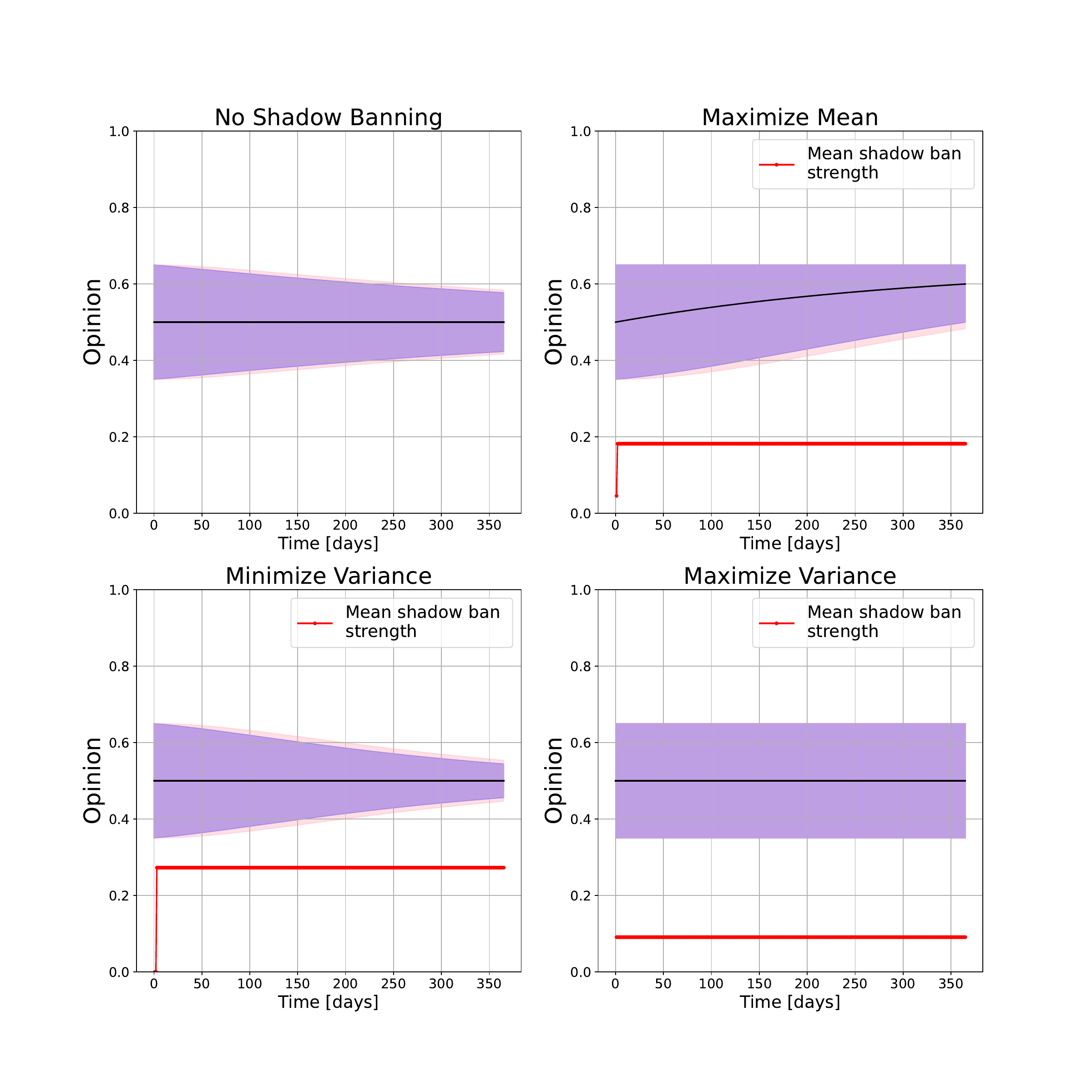}
    \caption{Opinion distributions and mean shadow ban strength versus time under shadow banning control policies for different objectives on a stochastic block model network.  For the opinions, the purple region is the 25th to 75th quantiles, and the pink region is the 5th to 95th quantiles.   The objectives are (top left) no shadow banning, (top right) maximize mean, (bottom left) minimize variance, and (bottom right) maximize variance.  }
    \label{fig:sbm_network_opinions}
\end{figure}

The shadow banning is applied in the same manner as with the path network, (daily update of policy with $s_{edge}=1$ and $s_{network}=0.5$).  We calculate the shadow banning policy for each objective function and show the resulting evolution of the opinions in Figure \ref{fig:sbm_network_opinions}.    With no shadow banning the network slowly approaches consensus at  0.5.  When maximizing the mean the shadow banning is able to push the opinions near 0.65, which is the maximum value in the initial opinions. Since shadow banning is only reducing content on the platform, the final opinions cannot be greater than the maximum opinion bounded by the initial values. Minimizing the variance causes the opinions to approach consensus, but  faster than without any shadow banning as can be seen by the narrower spread in the final opinion distribution.   When maximizing the variance, the opinions in each cluster stay at their initial values throughout the simulation.  Like with the path network, the mean shadow banning strength stays above zero for the simulation duration for each objective.  However, the value is lower than for the path network because fewer edges are shadow banned, as we will discuss next.

We visualize the early shadow banning policies for the stochastic block model network in Figure \ref{fig:sbm_network_policy} as was done for the path network.  The networks shown correspond to policies at $t=10$.  We did not use $t=0$ because the equality of the initial opinions within the clusters resulted in no shadow banning.  As the dynamics evolve the opinions take on different values and we obtain non-trivial shadow banning policies. For maximizing the mean, the shadow banning policy blocks the inter-cluster edges pointing from the lower opinion cluster to the higher opinion cluster.  This prevents the lower opinion cluster from pulling down the higher opinion cluster. Within the lower opinion cluster, edges pointing from the more extreme nodes to the boundary nodes are blocked to avoid these boundary connectors being pulled away from their higher opinion neighbors. Minimizing variance removes the edges pointing to the nodes on the cluster boundaries.  These edges pull the opinions to the extremes, so when trying to minimize the variance it is expected that they will be blocked. When maximizing the variance, the policy blocks all inter-cluster edges.  This is to be expected as those are the only edges that pull the opinions together.  In fact we see in Figure \ref{fig:sbm_network_opinions} that theses edges remain blocked for the entire simulation.  In contrast, the other two objective functions have the shadow banning turn off once the opinions reach consensus.

One important observation here is that shadow banning does not move the opinions beyond the maximum and minimum values in the initial condition.   Shadow banning cannot drive anyone to an extreme opinion unless those opinions already exist in the network.  This is in contrast to other methods to shift opinions which utilize bots that can drive opinions to arbitrary extremes \cite{hunter2022optimizing}.  The difference is that bots inject new content into the network which can have an extreme opinion.  Shadow banning can only remove content produced naturally in the network, so it cannot move anyone beyond the bounds determined by the users' initial opinions.

%%%%%%%%%%%%%%%%%%%%%%%%%%%%%%%%%%%%%%%%%%%%%%%%%%%%%%%
\subsection{Twitter Networks}
\subsubsection{Datasets}
We now apply shadow banning to a set of Twitter networks which have been utilized in previous studies on opinion dynamics \cite{des2022detecting, hunter2022optimizing}.  These datasets are ideal for us as they provide a network structure, posting rates, and opinions for a set of social media users engaged in an online conversation on politically polarizing topics.  The topics these networks cover are the 2016 United States presidential election and the Gilets Jaunes protests in France.  The raw datasets include tweets and also the follower graph formed by the users posting these tweets.  For each dataset, tweets were collected that contained specific keywords (a full list of keywords can be found in \cite{des2022detecting}).

Each user's posting rate was set equal to the number of their tweets in the dataset divided by the data collection period length.   The tweets' opinions on specific topics were measured using a neural network trained on a set of hand-labeled tweets.  The opinions were real numbers between zero and one.  For the U.S. election dataset an opinion of one represented a pro-Trump sentiment.  For the Gilets Jaunes dataset an opinion of one represented a pro-Gilets Jaunes sentiment.  Each user's opinion was calculated as the mean of the opinions of their tweets in the dataset.  Specific details on the collection and processing  of these datasets can be found in \cite{des2022detecting}.  We provide some summary statistics about the datasets in Table \ref{table:assess_data_stat}.  
\begin{table}[!hbt] \centering
	\caption{Basic information about the Twitter datasets. M is millions and K is thousands.}
	\label{table:assess_data_stat}
	\centering
	\begin{tabular}{|l|l|c|c|c|}
		\hline
		Event     &    Data collection   & Number of  & Number of & Number of\\
		&    period            &  tweets & follower edges & users\\
		\hline
		U.S. presidential  & 	Jan. 2016 - & 2.4M & 5.4M & 78K \\ 
        election  & Nov 2016 &  & &  \\\hline
		Gilets Jaunes  & 	Jan. 2019 -  & 2.3M & 4.6M & 40K \\
                       & Apr. 2019 &  & &  \\\hline
	\end{tabular}
\end{table}  

The U.S. election dataset consists of tweets by Twitter users who posted about the second debate of the 2016 U.S. presidential election  between Hillary Clinton and Donald Trump. This dataset has 2.4 million tweets posted by 77,563 users.  The resulting follower graph contained 5.4 million edges. The Gilets Jaunes, also known as the Yellow Vests movement, emerged in France in November 2018. Initially sparked by a significant hike in fuel prices, it rapidly expanded into a widespread protest against the policies of President Emmanuel Macron's government.   The Gilets Jaunes dataset consists of tweets between January 26th, 2019 to April 29th, 2019 that contained Gilets Jaunes related keywords. The resulting dataset contained 2.3 million tweets, 40,456 users, and 4.6 million edges in the associated follower graph.

For our simulations we use the subgraph induced by a random subset of users for each dataset. Each subgraph has 15,000 users with opinions less than or equal to 0.5 and 15,000 users with opinions greater than 0.5. For the U.S. election dataset, the resulting sampled network has 30,000 users and 844,563 edges.  The Gilets Jaunes sampled network has 30,000 users and 1,084,678 edges. The network sizes are chosen to resemble the size of the networks used in a field experiment and observational study concerning content moderation.  The field study in \cite{Nyhan2023} recruited 23,377 US-based adult Facebook users to assess the impact of modifying the polarity of content seen by users on their political polarization.  The observational study in \cite{jaidka2023silenced} audited a random sample of 25,000 Twitter accounts to identify if they were shadow banned. In addition to replicating the size of networks in these works, using a subgraph of our data also reduces the computational time of the simulations. 

%%%%%%%%%%%%%%%%%%%%%%%%%%%%%%%%%%%%%%%%%%%%%%%%%%%%%
\subsubsection{Simulation Results}
Our shadow banning simulations have a similar form to that for the synthetic networks.  Shadow banning policies are calculated daily, with the maximum mean shadow banning strength $s_{network}$ set to 0.05, and no limit to the shadow banning strength on each individual edge ($s_{edge}=1$). The maximum mean shadow banning strength of 5\% is chosen based on \cite{jaidka2023silenced} that found 6.2\% of sampled Twitter accounts were shadow banned at least once within a year of data collection. In addition, \cite{LeMerrer2021} estimated that between  0.5\% to 2.3\% of users were banned in the Twitter networks they studied. We do not limit $s_{edge}$ as shadow banning usually lasts at least 24 hours and up to two weeks on commonly used social media platforms, and we update our control daily. Our simulations cover 365 days.  The opinion dynamics model is the bounded confidence model with $\epsilon = 0.1$ and $\omega = 0.003$.  We also repeat our analysis on variations of these model parameters (see Appendix).

\begin{figure}[t]
    \centering
    \includegraphics[width=\textwidth]{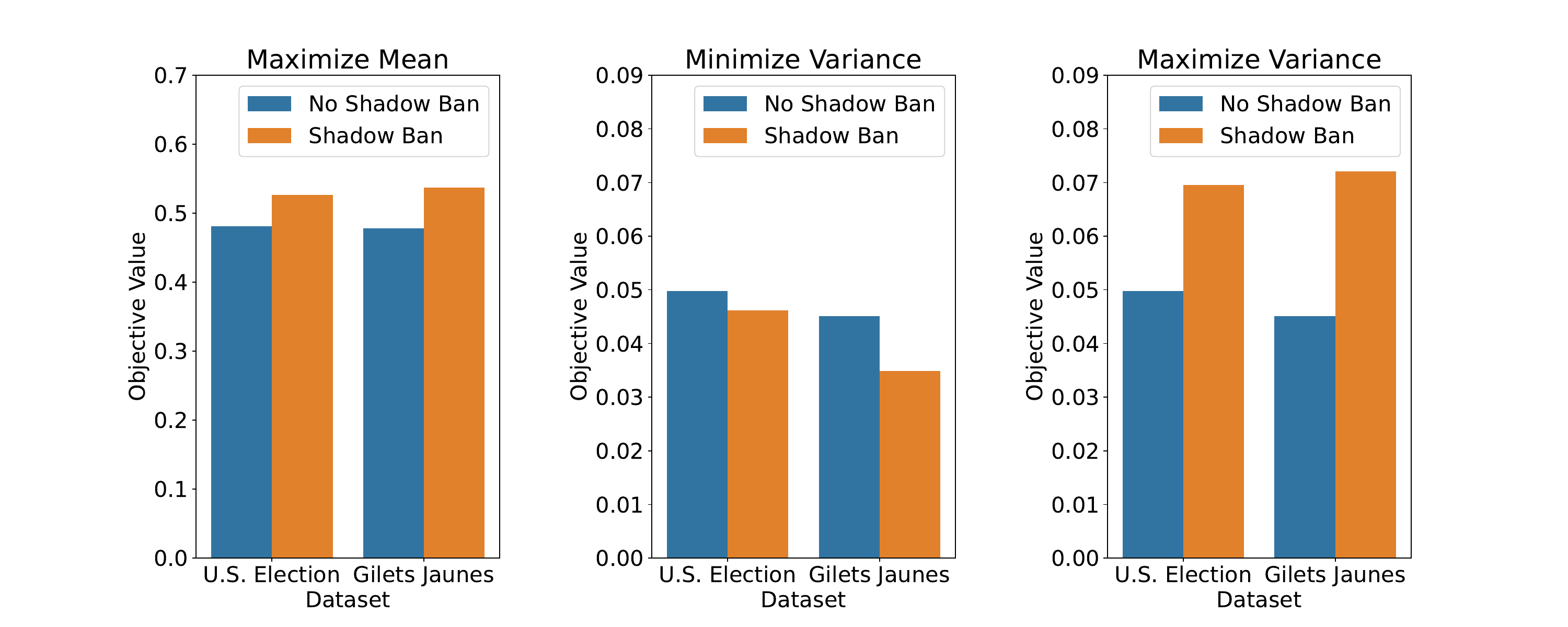}
    \caption{Bar plots of terminal objective values with (blue) no shadow banning versus (orange) shadow banning for the U.S. election and Gilets Jaunes datasets, with objectives being (left) maximize mean, (middle) minimize variance, and (right) maximize variance. For the variance objectives, the terminal variances are reported. The objective improvements by shadow banning compared to no shadow banning are (by U.S. election and Gilets Jaunes) 9\% and 12\% for maximizing mean, 7\% and 23\% for minimizing variance, and 40\% and 60\% for maximizing variance.}
    \label{fig:objectives_comparison}
\end{figure}

We plot the terminal objective value in the simulation for each dataset in Figure \ref{fig:objectives_comparison}. As can be seen, the shadow banning policy is able to improve the objective value relative to no shadow banning by 7\% to 60\% depending on dataset and objective.  We see that the shadow banning policy is able to shift the opinion mean, decrease the variance, and also increase the variance.  Therefore, we see the variety of opinion manipulations we can achieve with shadow banning, even with limited mean shadow banning strength. Next we explore the evolution of the opinions in more details to understand how the shadow banning is affecting the opinions.

\begin{figure}[H]
    \centering
    \includegraphics[width=\textwidth]{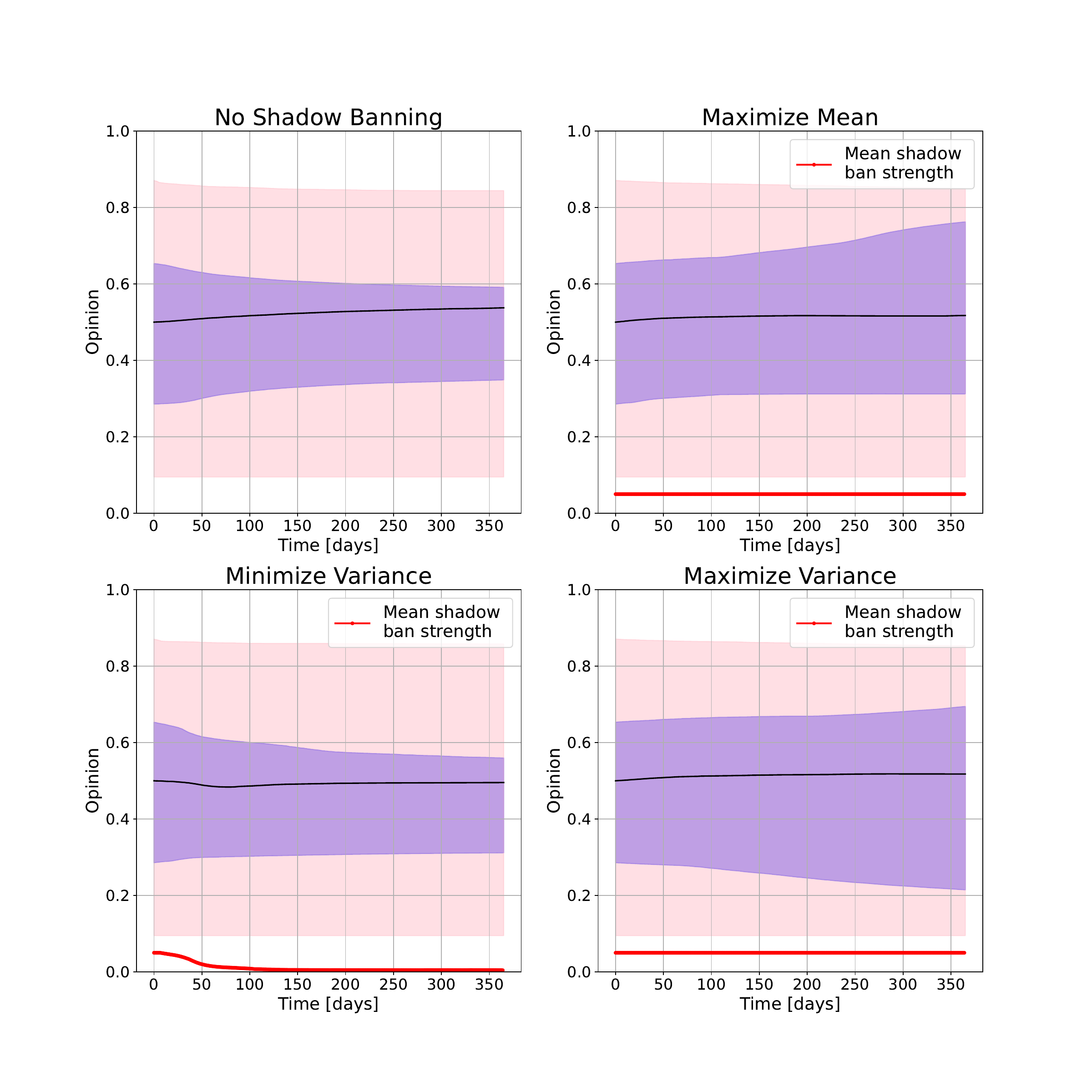}
    \caption{Opinion distributions and mean shadow ban strength versus time under shadow banning control policies for different objectives on the 2016 U.S. presidential election Twitter network.  For the opinions, the purple region is the 25th to 75th quantiles, and the pink region is the 5th to 95th quantiles.  The objectives are (top left) no shadow banning, (top right) maximize mean, (bottom left) minimize variance, and (bottom right) maximize variance.}
    \label{fig:uselection_opinions}
\end{figure}

\begin{figure}[H]
    \centering
    \includegraphics[width=\textwidth]{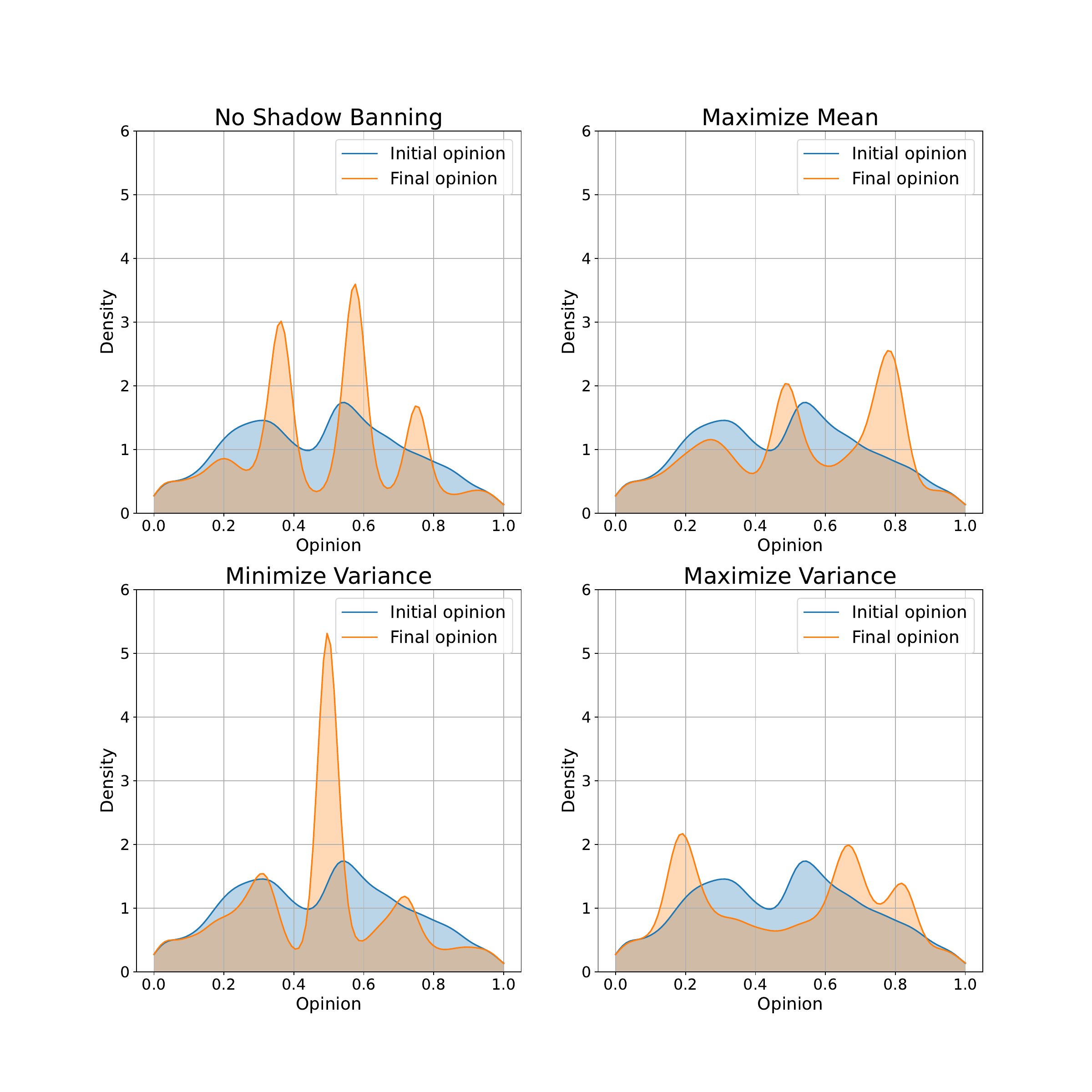}
    \caption{Initial and final opinion densities under shadow banning control policies for different objectives on the  2016 U.S. presidential  election  Twitter network. The objectives are (top left) no shadow banning, (top right) maximize mean, (bottom left) minimize variance, and (bottom right) maximize variance.}
    \label{fig:uselection_density}
\end{figure}

We begin with the U.S. presidential election dataset.  We show the opinion evolution under no shadow banning and with shadow banning for different objectives in Figure \ref{fig:uselection_opinions}.  We also show density plots of the initial and final opinions in Figure \ref{fig:uselection_density}.  Our first observation is that over the one year simulation the opinion quantiles show a very small movement.  This is due to our bounded confidence model specification as we would expect social media users not to experience a major change in opinion over this time period.  While the changes are small, they differ significantly depending on the shadow banning objective.  With no shadow banning, the opinions show a slight movement towards the center. The density plot shows that opinions are converging around significant values present in the initial distribution, resulting in three major modes: left, center, and right. This pattern closely resembles real-world election polls. When maximizing the mean, we see that the 75th quantile is driven upwards from an initial value of 0.65 to a final value of 0.75.  Looking at the density plots, we see that the increase in the upper quantiles is primarily due to the creation of a mode centered around 0.8. Minimizing the variance does not impact the median opinion, but slightly pulls in the 25th and 75th quantiles.  From the density plot we see that the shadow banning has pulled the opinions towards the center at 0.5.  Maximizing the variance appears to widen the 25th and 75th quantiles over time.  The density plot shows that this policy has removed opinions from the center.

The mean shadow banning strength shows different behavior for the different objectives.  For minimizing variance, the shadow banning strength decays to zero.  For maximizing the mean and variance, it remains at maximum strength over time. The opinion dynamics are attractive, so less shadow banning is needed for minimizing variance as the natural dynamics assists in driving the variance towards zero. However, maximizing variance requires driving opinions apart, which goes against the natural opinion dynamics, and so constant shadow banning is needed.  We see the same behavior for maximizing the mean, and this is due to the initial opinion distribution in the network not having a large proportion of users with high opinions.  The constant shadow banning is needed so that these users are not pulled down and can continuously pull other users up.

\begin{figure}[H]
    \centering
    \includegraphics[width=\textwidth]{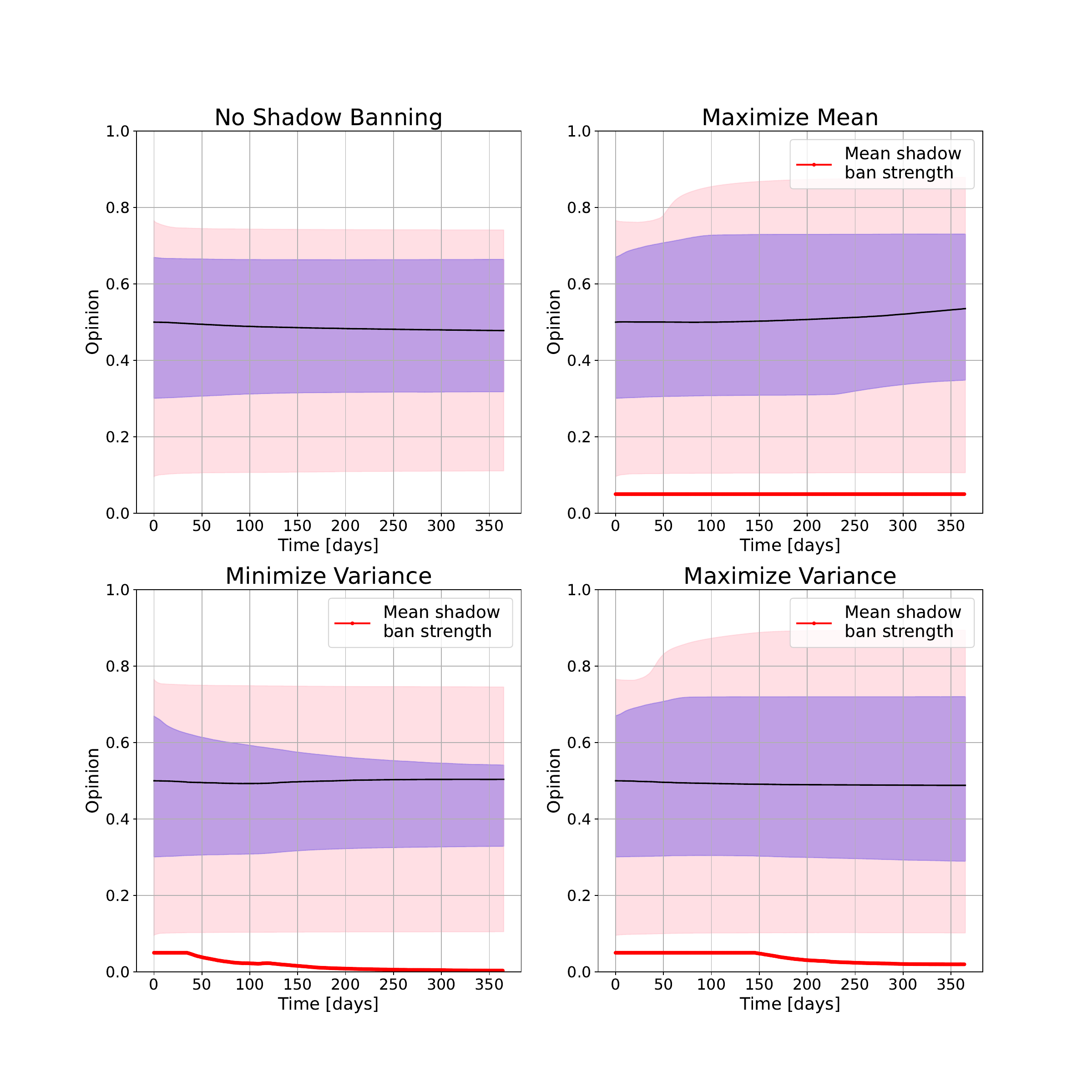}
    \caption{Opinion distributions and mean shadow ban strength versus time under shadow banning control policies for different objectives on the Gilets Jaunes Twitter network.  For the opinions, the purple region is the 25th to 75th quantiles, and the pink region is the 5th to 95th quantiles. The objectives are (top left) no shadow banning, (top right) maximize mean, (bottom left) minimize variance, and (bottom right) maximize variance.}
    \label{fig:giletsjaunes_opinions}
\end{figure}

\begin{figure}[H]
    \centering
    \includegraphics[width=\textwidth]{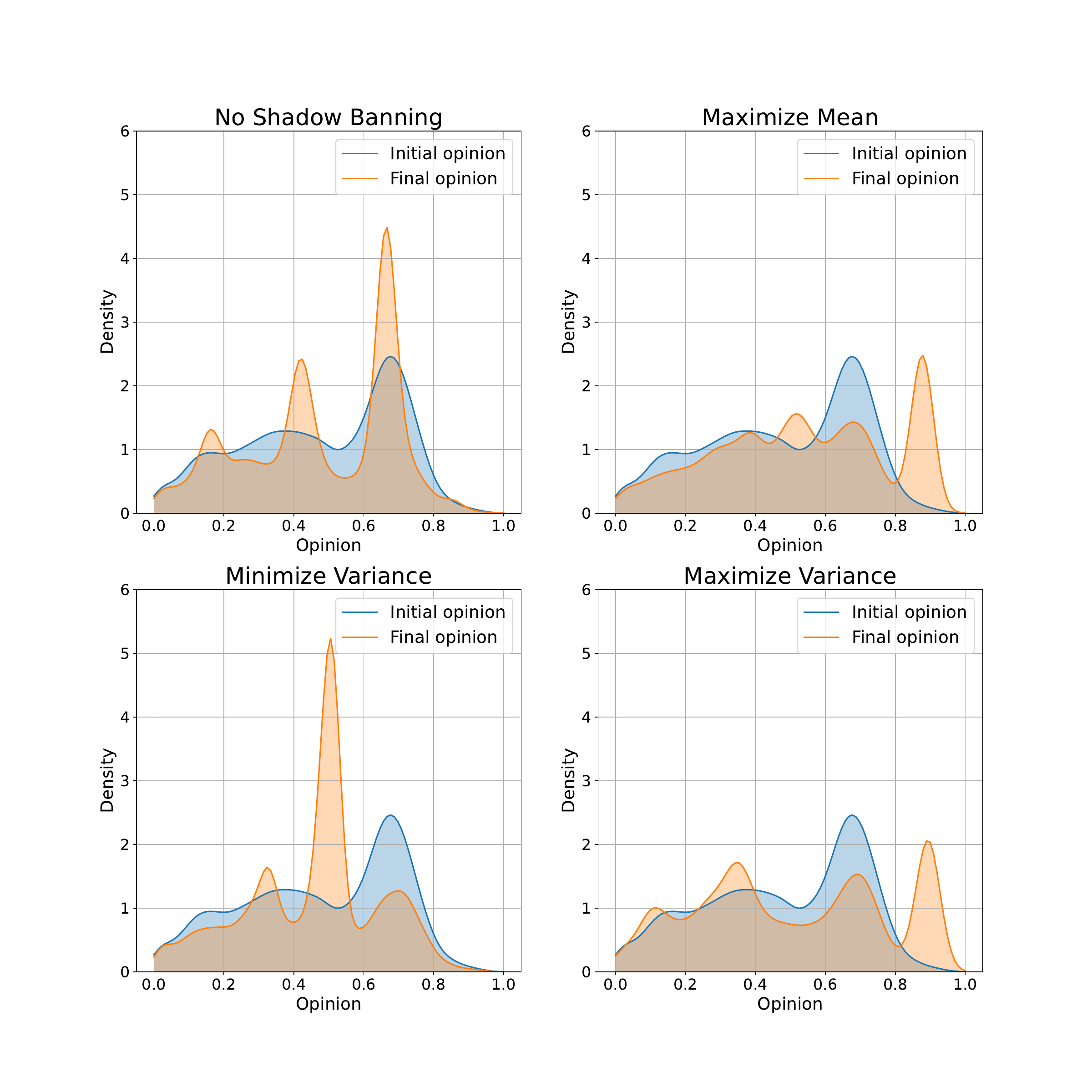}
    \caption{Initial and final opinion densities under shadow banning control policies for different objectives on the Gilets Jaunes Twitter network.  The objectives are (top left) no shadow banning, (top right) maximize mean, (bottom left) minimize variance, and (bottom right) maximize variance.}
    \label{fig:giletsjaunes_density}
\end{figure}

We next look at the Gilets Jaunes Twitter network, with the opinion evolutions shown in Figure \ref{fig:giletsjaunes_opinions} and initial and final opinion densities shown in Figure \ref{fig:giletsjaunes_density}.  The natural dynamics of the network do not appear to move the opinion quantiles much. Maximizing the mean and variance result in similar final opinion distributions.  The difference between the final 25th and 75ht quantiles is large for both objectives, but slightly larger for maximizing the variance.  The final median is slightly higher for maximizing the mean.  Apart from these differences, we find that maximizing either objective results in a network with a slightly increased opinion median and highly polarized opinions.  Looking at the final opinion densities in Figure \ref{fig:giletsjaunes_density} we see that they are nearly the same for the two objectives.  Minimizing the variance results in the opinion becoming concentrated at the center, as can be seen by the decrease in separation between the 25th and 75th quantiles in Figure \ref{fig:giletsjaunes_opinions}.  From the density plot we see that the shadow banning policy is creating a mode at 0.5 with a narrow width.

The mean shadow ban strength has a similar behavior for the Gilets Jaunes network as for the U.S. presidential debate network.  Minimizing the variance requires less shadow banning as the natural dynamics assist in creating consensus.  Maximizing the mean and variance require more shadow banning. The difference  is that the shadow banning strength for maximizing the variance decreases slowly over time as the opinions reach the extreme ends of the spectrum. The reason for the shadow banning decay is that once the opinions are away from the middle, then the natural attractive opinion dynamics takes over, pulling the opinions towards the extremes. %Although not displayed here, we find that the assist of the natural opinion dynamics also occurs when maximizing the opinion mean under opinions dynamics with larger values for $\epsilon$ and $\omega$. Once opinions reach the upper extreme, the shadow banning strength weakens. 

Our simulation studies provide valuable insights into the characteristics of shadow banning. Firstly, shadow banning proves to be a versatile tool for influencing opinions, with the potential to produce a range of effects, including steering opinions in specific directions, moderating polarization, or amplifying it. Secondly, the initial distribution of opinions within the network emerges as a significant determinant in shaping the shadow banning strategy. Even when pursuing the same objective, the resulting policies and trajectories of opinion evolution can exhibit substantial variations based on the network's structure and the initial opinion distribution. Lastly, shadow banning demonstrates a notable degree of adaptability. Its necessary intensity fluctuates over time, depending on the evolving state of the network. In certain scenarios, as the network naturally progresses toward a desired state, the need for intense shadow banning diminishes. Conversely, in other cases, continuous shadow banning is required for maintaining the desired opinion trajectory.
%%%%%%%%%%%%%%%%%%%%%%%%%%%%%%%%%%%%%%%%%%%%%%%%%
\subsubsection{Partisan Bias in Shadow Banning}

One can choose an objective with a partisan bias when shadow banning.  For instance, one can make the objective function be the mean (or negative mean) if one wants to shift the opinions up (or down).  This is a clearly biased objective favoring one extreme of a topic.  However, the implemented policy will not appear overtly partisan.  To measure the overt partisan nature of a shadow banning policy at any given point in time, we segment users into two political groups based on their current opinion.  For the U.S. presidential election dataset, we label Democrats as those with opinion less than or equal to 0.5, and Republicans as those with opinion greater than 0.5. For the Gilets Jaunes dataset, Gilets Jaunes opponents have opinion less or equal to 0.5, and Gilets Jaunes supporters have opinion greater than 0.5.  We then look at the fraction of users shadow banned in each political group at a given time, which we refer to as the \emph{shadow ban rate}.  A user $i$ is considered shadow banned at time $t$ if the shadow ban strength $u_{ij}(t)$ is greater than zero for at least one $j$.  This means at least one follower of $i$ is not seeing all content posted by $i$.  

\begin{figure}[ht]
    \centering
    \includegraphics[width=\textwidth]{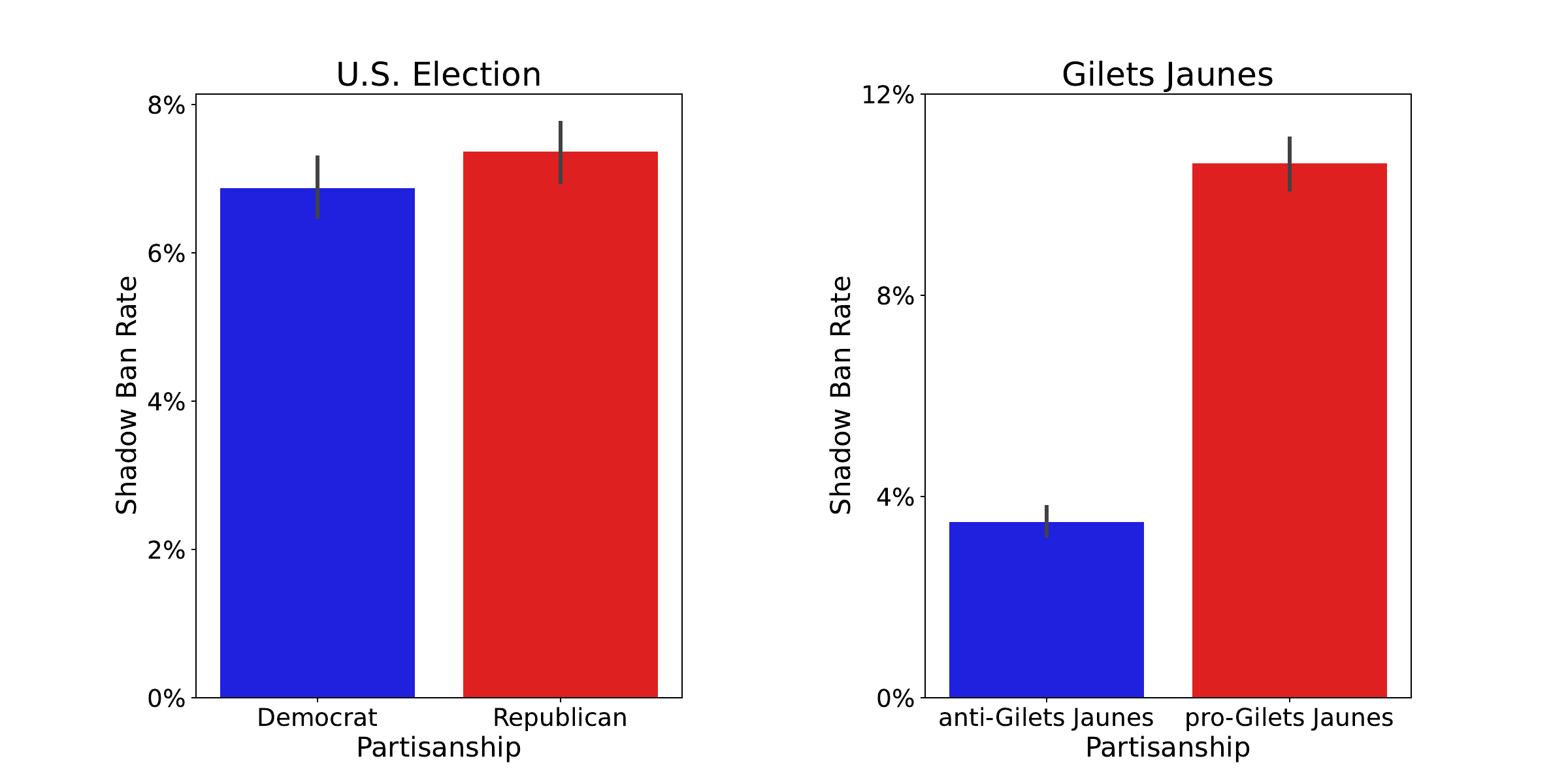}
    \caption{Bar plots of shadow ban rates by partisan group at $t=0$ for the (left) U.S. election and (right) Gilets Jaunes datasets.  The shadow banning objective is to maximize the opinion mean.  For the U.S. election, this means to shift the mean towards Republicans.  For Gilets Jaunes this means to shift the opinions towards pro-Gilets Jaunes. Shadow ban rate here is calculated by the fraction of number of accounts, or vertices, that have at least one out-degree edge that is shadow banned. Error bars indicate the 95\% confidence interval of the mean estimate.
    }
    \label{fig:partisan}
\end{figure}

We would expect a political bias in the shadow ban rates given that the objective is to maximize the opinion mean.  However, we find this is not the case.  We plot the shadow ban rate at the initial time ($t=0$) in our simulations in Figure \ref{fig:partisan}.  For the U.S. presidential election dataset, the two values are nearly identical, with the Republicans being shadow banned at a slightly higher rate than the Democrats.  A more extreme result is found for Gilets Jaunes.  We see that the pro-Gilets Jaunes users are shadow banned at nearly three times the rate of the anti-Gilets Jaunes users.  These findings are counter-intuitive as they indicate that the shadow banning policies have a bias that is opposite the bias of the objective.  However, from the opinion evolution plots in Figures \ref{fig:uselection_opinions} and \ref{fig:giletsjaunes_opinions}, we see that these policies lead to opinions distributions which exhibit the bias suggested by the objective. 

To understand why a shadow banning policy can appear unbiased while being very biased, it is useful to consider again the path network discussed earlier.  The initial shadow banning policy for maximizing the mean is shown in Figure \ref{fig:linear_network_policy}.  From the figure, we see that every node is shadow banned except for the red node with the highest opinion located at the end of the path. Specifically, the edge pointing to the neighbor with higher opinion is shadow banned.  The remaining edges indicate that the posts can flow from nodes with higher opinion to those with lower opinion.  This has the effect of  only allowing upward opinion shifts, which causes the opinion mean to increase over time.  However, every node (except for the maximum opinion node) has a neighbor with higher opinion.  This means that all of these nodes are shadow banned, which causes  the policy to appear unbiased.  

In general, for maximizing the opinion mean, the shadow banning policy blocks any edge which pulls opinions downwards.  These edges can be incident on nodes of either partisan group.  In this manner the policy appears unbiased, or even possibly biased in the opposite direction depending upon the network structure and opinion distribution.  The natural approach is to consider which users are shadow banned, which would allow biased shadow banning to go undetected.  Our results suggest that to measure a bias in a shadow banning policy, one must look at the edges which are shadow banned, and not the users.  In particular, one must look at the sign of the opinion shift among the shadow banned edges to identify the bias.

Our result shows the danger of shadow banning.  One would think that if a social media platform employed an overtly biased content moderation policy, this bias would be easily observed.  However, we find that the platform can employ a shadow banning policy which appears to be unbiased, yet over time creates a bias in the users' opinions.  The platform's efforts at shifting opinions would likely go undetected as the actual implemented policy seems unbiased, or even biased in the opposite direction.  One would not realize there was a bias in the policy until after it has been employed for a long period of time.
%%%%%%%%%%%%%%%%%%%%%%%%%%%%%%%%%%%%%%%%
\subsubsection{Sensitivity Analysis}
We investigate the sensitivity of the performance of shadow banning policies as a function of the maximum mean shadow ban strength $s_{network}$ and edge shadow ban strength $s_{edge}$. We investigate the sensitivity with respect to the opinion dynamics model parameters in Appendix.

\begin{figure}[H]
    \centering
    \includegraphics[width = \textwidth]{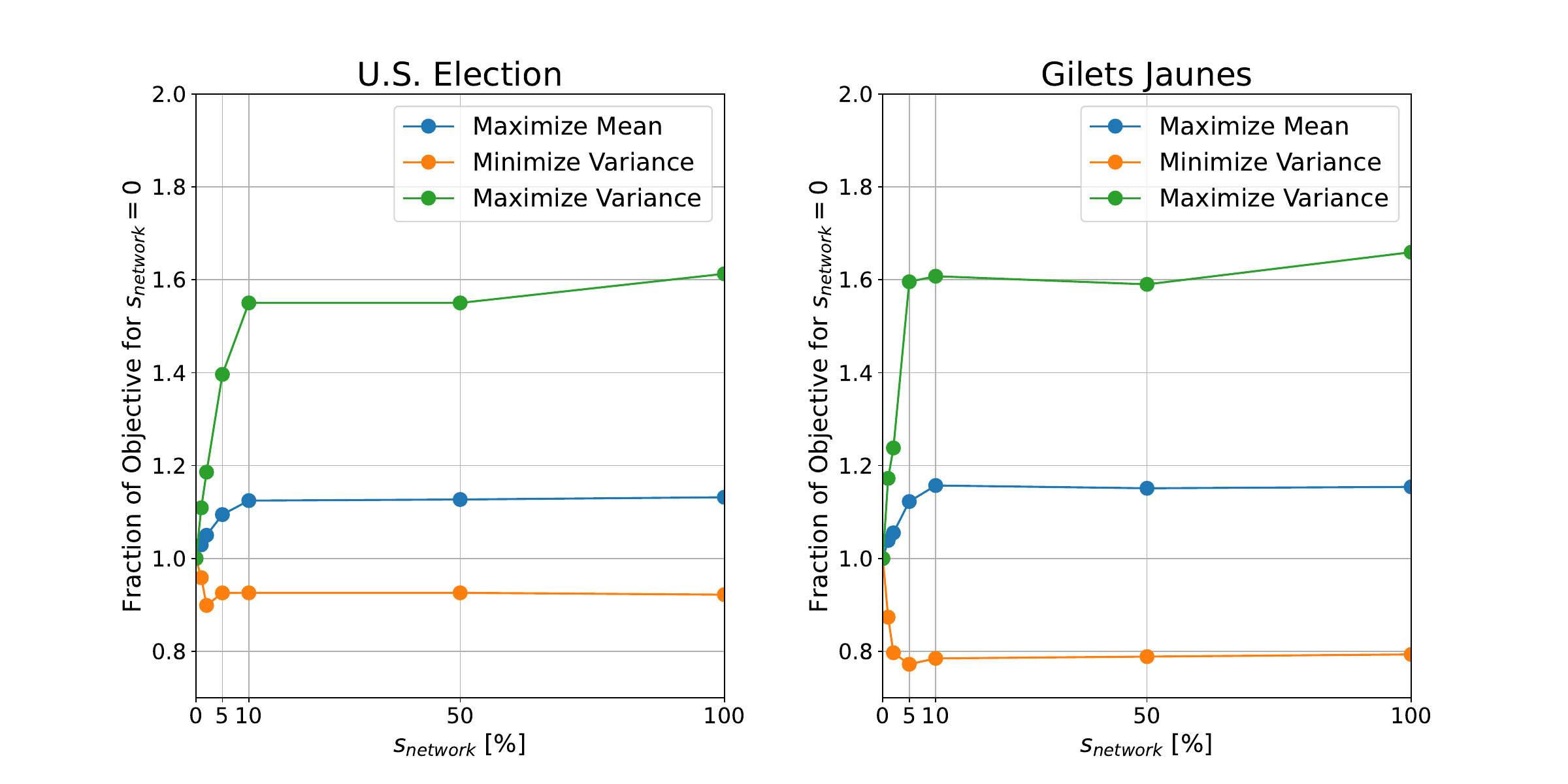}
     \caption{The terminal objective values as a function of $s_{network}$ for the  (left) U.S. election and (right) Gilets Jaunes datasets ($s_{edge}=1$). The y-axis shows the relative magnitude of the objective value compared to that of no shadow banning.}
    \label{fig:smax_sens}
\end{figure}

We first see how performance changes if we vary $s_{network}$ with $s_{edge}=1$. We consider the terminal value of the objective over the duration of the simulation.  Figure \ref{fig:smax_sens} shows the relative change of the different objectives  in the simulation relative to no shadow banning as $s_{network}$ is increased for the two datasets.  We find that the objectives plateaus for values of $s_{network}$ greater than 10\% for both networks. There appears to be no benefit to applying stronger global shadow banning beyond this value. This most likely occurs because the shadow banning is applied to a limited number of critical edges at each time step. Therefore, the opinions can be shifted without shadow banning a significant fraction of the edges.  

\begin{figure}[H]
    \centering
    \includegraphics[scale=0.5]{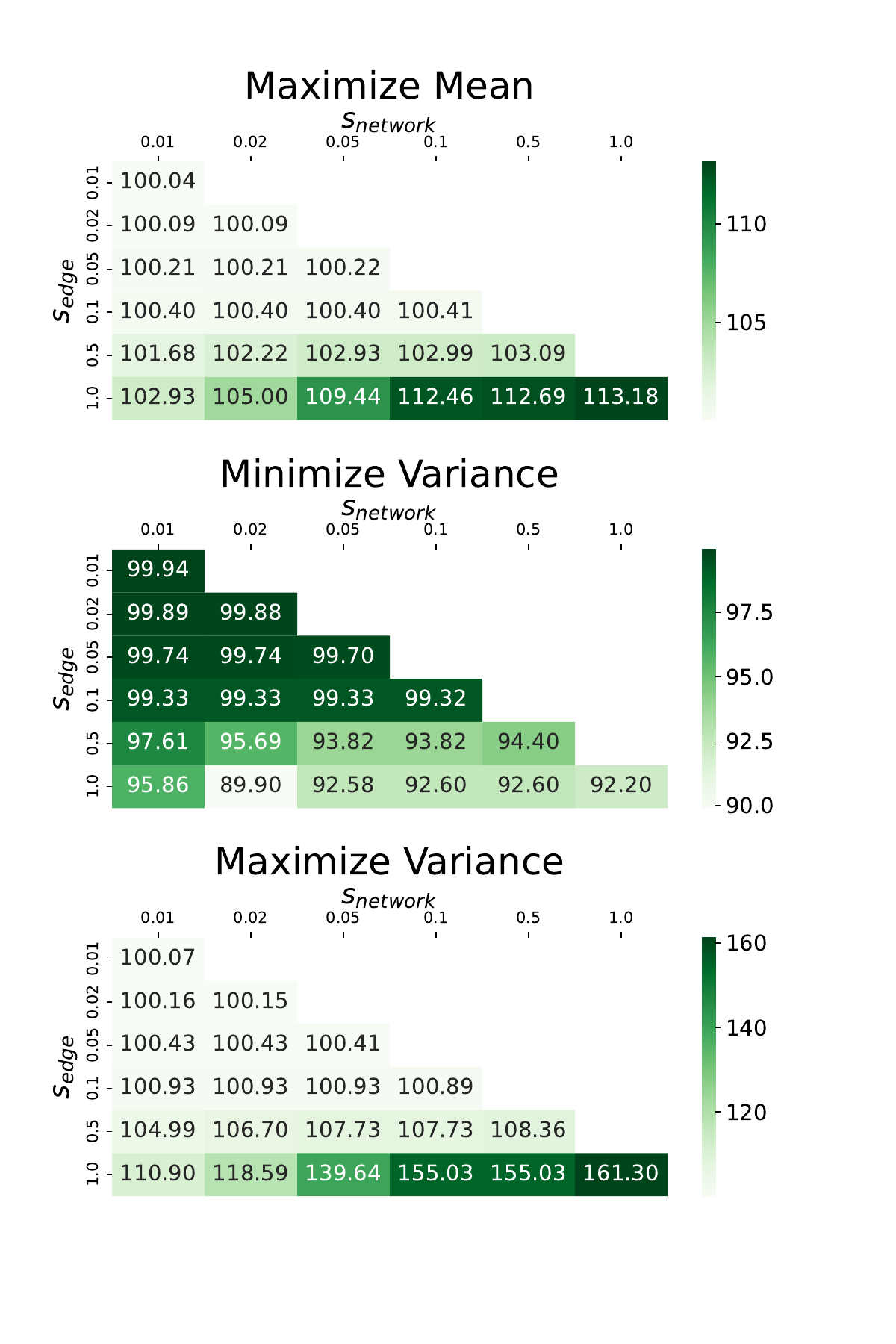}
     \caption{Terminal objective values for the U.S. election dataset as a function of $s_{network}$ and $s_{edge}$. The objectives are (top) maximize mean, (middle) minimize variance, and (bottom) maximize variance. Values in the cells are the magnitude relative to no shadow banning in percent.}
    \label{fig:sens_sbstrength_USE}
\end{figure}

\begin{figure}[H]
    \centering
    \includegraphics[scale=0.5]{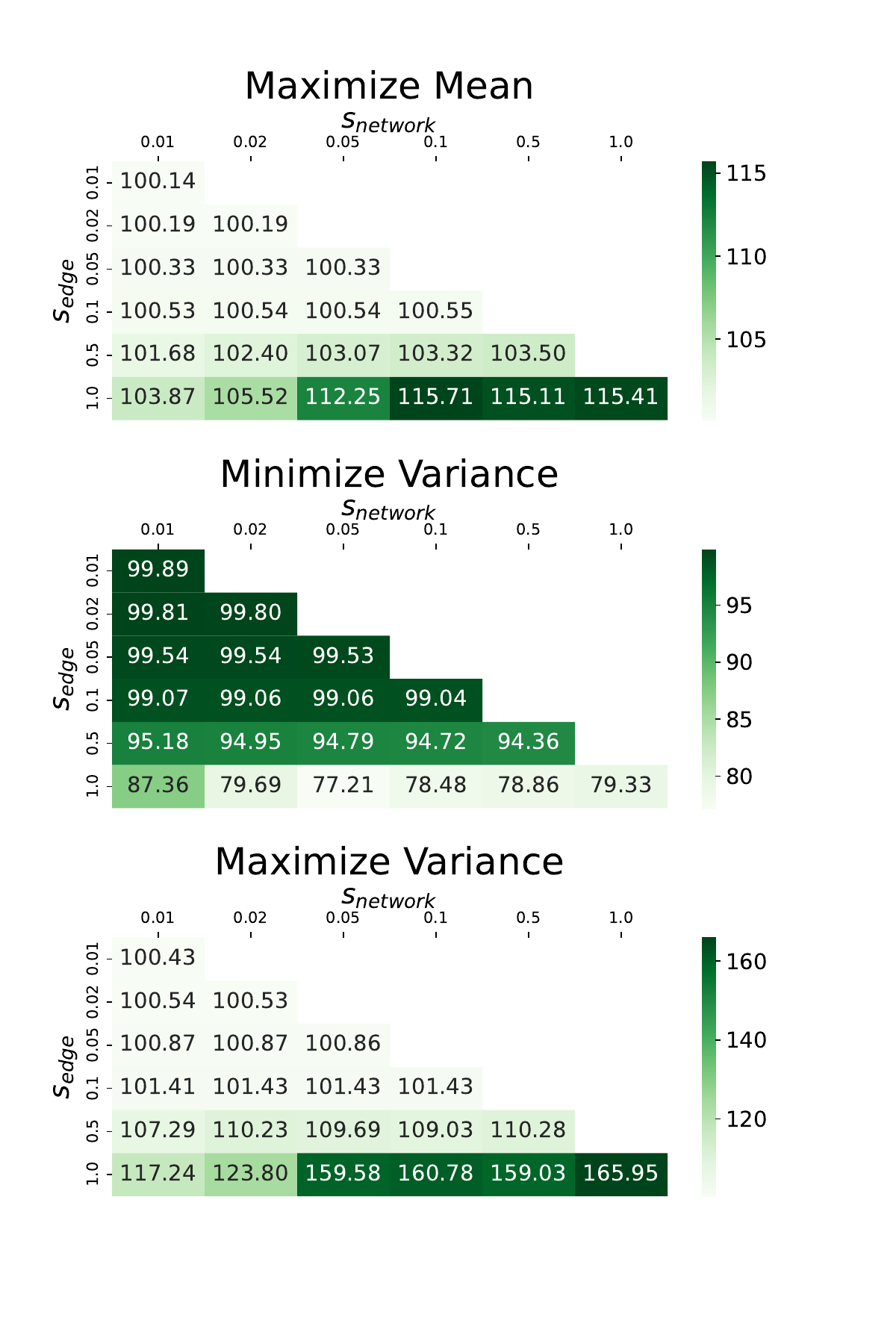}
 \caption{Terminal objective values for the Gilets Jaunes dataset as a function of $s_{network}$ and $s_{edge}$.  The objectives are (top) maximize mean, (middle) minimize variance, and (bottom) maximize variance.  Values in the cells are the magnitude relative to no shadow banning in percent.}
    \label{fig:sens_sbstrength_GJ}
\end{figure}

We next investigate the impact of $s_{edge}$ on performance.  We provide plots of how the terminal objective changes with respect to both $s_{edge}$ and $s_{network}$ for each dataset in Figures \ref{fig:sens_sbstrength_USE} and \ref{fig:sens_sbstrength_GJ}.  We find that for values of $s_{edge}$ less than 0.5 there is very little change in the objective relative to no shadow banning.  For larger values of $s_{edge}$ we see the shadow banning causing a non-trivial change in the objective values.  This shows that strong shadow banning needs to be allowed on the targeted edges in order to produce a non-trivial shift in the opinion distribution.  Therefore, while the shadow banning strength can be very low across the network, the critical edges that are targeted require a substantial amount of shadow banning.

%Note that with objective being maximizing the variance, we observe an $\epsilon$ threshold that impedes the objective. Our shadow banning policy could not maximize variance if $\epsilon$ is greater than 0.3 and 0.5, for the two datasets respectively. This is because the shift function is assumed attractive, and thus offsetting the polarization efforts from shadow banning. We also see that shadow ban strength remains at 0.05 all the time, showing that greater efforts are needed to diverge the opinions than to reach a consensus. Although in our setting the shift parameters are measured and not controls, future research could study how to influence $\epsilon$ to counter intended polarization efforts and to strengthen network resiliency.

%%%%%%%%%%%%%%%%%%%%%%%%%%%%%%%%%%%%%%%%%%%%%%%%%%%%%%%%%%%%
\section{Discussion}
Our findings show the power and flexibility of shadow banning as a content moderation tool for online social media platforms.  Precise shadow banning policies can be easily calculated for large networks by solving a linear program.  By applying these shadow banning policies, platforms can exert delicate influence over the distribution of user opinions. While this can serve goals like reducing polarization or curbing misinformation, it also holds the potential for misuse.  Shadow banning can intensify polarization within a network.  Platforms might use shadow banning to steer opinions towards or away from specific topics. Additionally, platforms might employ this biased shadow banning and remaining unnoticed due to the outward appearance of political neutrality.

The danger of a social media platform engaging in biased shadow banning is significant.  The effects are slow, and the bias can be undetected.  Over time, this can lead to dangerous outcomes for which it is too late to prevent.  Election outcomes can potentially be changed by such manipulation.  Societies can be polarized to the point of instability.  Intelligent policies should be enacted to prevent such abuse by social media platforms.  Conventional measures such as shadow ban rates may not reveal the bias exerted by the platforms.  However, more precise measures, such as shadow ban rates for edges of different opinion shift polarity, can reveal this bias.  Such measures should be employed to ensure that social media platforms use shadow banning to maintain platform health and safety and not for other malicious purposes.

Our research has a framework for investigating how content moderation, particularly in the context of shadow banning, impacts user opinions. This framework can also be expanded to analyze the influence of content recommendation algorithms on opinions. An intriguing path for future research is to utilize our framework to evaluate the effects of content recommendation algorithms on opinion polarization. This avenue of exploration could pave the way for designing content recommendation algorithms that not only enhance user experiences but also proactively address the potential for increased polarization.

\newpage
\section{Appendix}

%%%%%%%%%%%%%%%%%%%%%%%%%%%%%%%%%%%%%%%%%%%%%%%%%%%%%%%%%%
%\subsection{Robustness for Variations of Shadow Ban Strength Constraint}
%Sensitivity of the objectives as a function of the shadow ban strength constraints including network average $s_{network}$ and individual edge $s_{edge}$ is illustrated in the heat maps in Figures \ref{fig:sens_sbstrength_USE} and \ref{fig:sens_sbstrength_GJ} for the two datasets respectively. All combinations of $s_{network} \in [0.01, 0.02, 0.05, 0.1, 0.5, 1]$ and $s_{edge} \in [0.01, 0.02, 0.05, 0.1, 0.5, 1]$ are simulated and time-average rewards compared to no shadow banning are reported in the cells. Here we fix the shift parameters where $\epsilon=0.1$ and $\omega=0.003$. The upper triangles of the heat maps are masked since the results of $s_{edge} < s_{network}$ are the same as those of $s_{edge} = s_{network}$, as maximum strength on individual edge is a more strict constraint for controls than the mean strength limit on the entire network.

%Compared to no shadow ban where $s_{network}=0$ and $s_{edge}=0$, we see that for maximizing the mean and variance, our shadow ban policy gives larger time-average rewards, regardless of the choice for $s_{network}$ and $s_{edge}$. For minimizing the variance, the policy leads to lower time-average variances for all combinations of the shadow ban strength limits. Thus, our shadow banning policy is robust for variations of maximum shadow banning strength.

%%%%%%%%%%%%%%%%%%%%%%%%%%%%%%%%%%%%%%%%%%%%%%%%%%%%%%%%%%%%%%%%%%%%%%
\subsection{Robustness for Variations of the Bounded Confidence Model}
\begin{figure}
    \centering
    \includegraphics[scale=0.5]{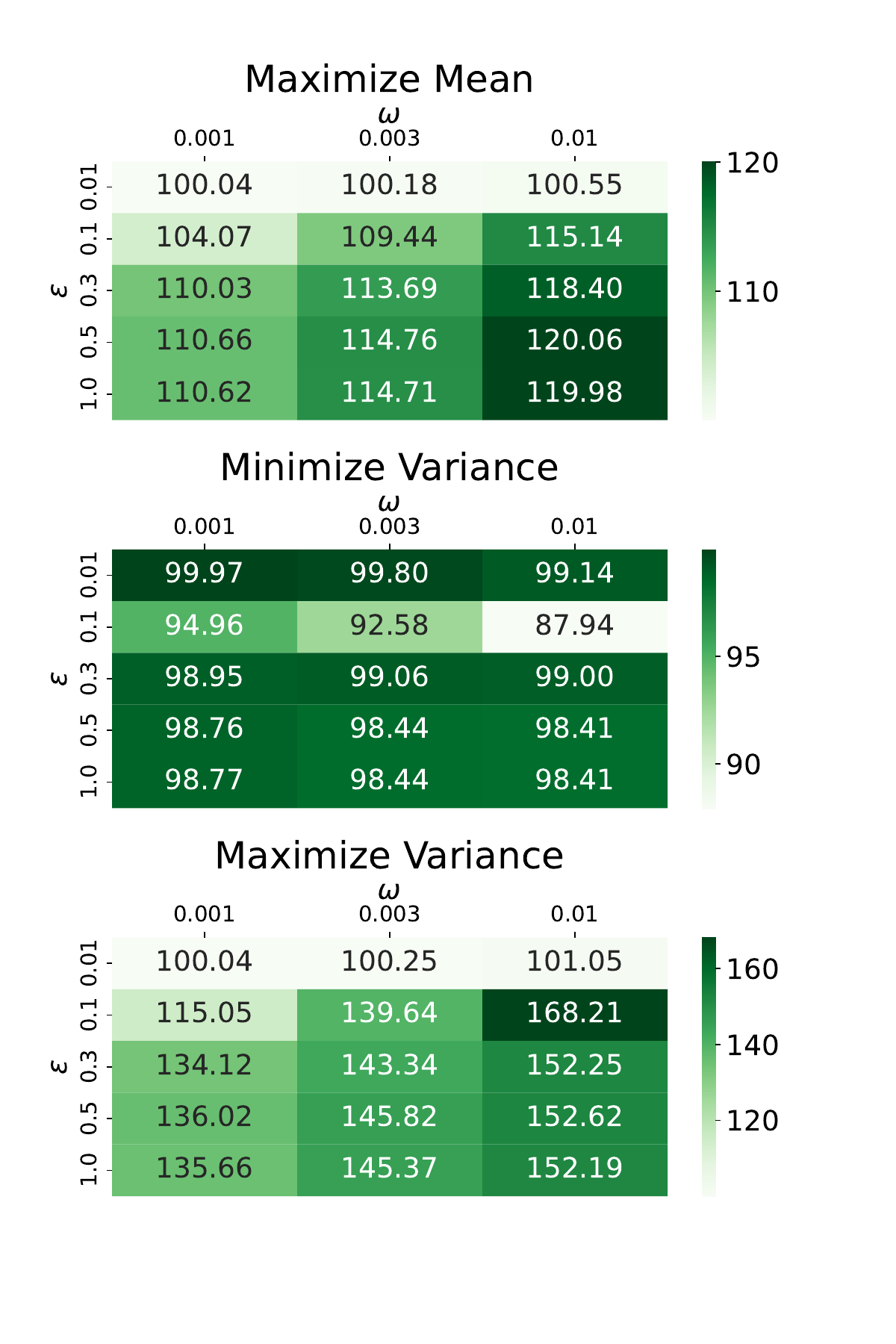}
    \caption{Terminal objective values for the U.S. election dataset as a function of $\epsilon$ and $\omega$.  The objectives are (top) maximize mean, (middle) minimize variance, and (bottom) maximize variance. Values in the cells are the magnitude relative to no shadow banning in percent.
    }
    \label{fig:sens_shift_USE}
\end{figure}

\begin{figure}
    \centering
    \includegraphics[scale=0.5]{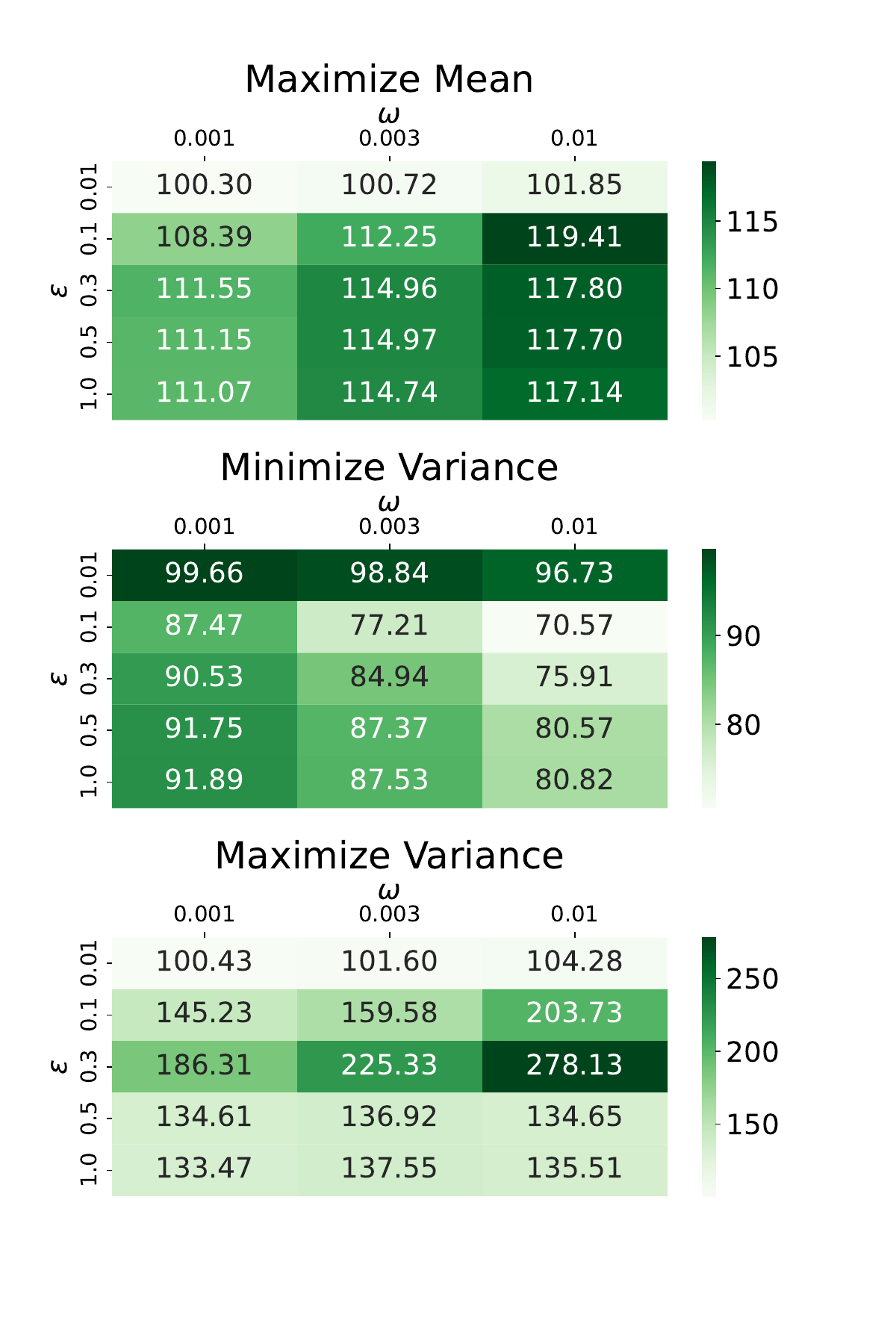}
    \caption{Terminal objective values for the Gilets Jaunes dataset as a function of $\epsilon$ and $\omega$.  The objectives are (top) maximize mean, (middle) minimize variance, and (bottom) maximize variance. Values in the cells are the magnitude relative to no shadow banning in percent.}
    \label{fig:sens_shift_GJ}
\end{figure}

We present here the performance  of different shadow banning objectives on the U.S. presidential election and Gilets Jaunes Twitter networks under different specifications of the bounded confidence model. All combinations of $\epsilon \in [0.01, 0.1, 0.3, 0.5, 1]$ and $\omega \in [0.001, 0.003, 0.01]$ are simulated and the terminal objective values compared to no shadow banning are illustrated in the heat maps in Figures \ref{fig:sens_shift_USE} and \ref{fig:sens_shift_GJ} for each dataset. Shadow banning strength limits are fixed with $s_{network}=0.05$ and $s_{edge}=1$.

Our first observation is that regardless of the choice of $\epsilon$ and $\omega$ in the bounded confidence model, our policy leads to an improvement in the objective relative to no shadow banning.  This shows that our shadow banning policies show some level of robustness with respect to the bounded confidence model.

For maximizing the mean, the objective smoothly increases as the persuasion strength is increased. However, the variance objectives show some more interesting behavior. In the U.S. election dataset, when minimizing the variance, larger $\epsilon$ values do not offer as much improvement as $\epsilon=0.1$. This is because stronger opinion dynamics play a more dominant role in determining the location of opinion consensus, overshadowing the impact of shadow banning. When maximizing the variance, the most substantial terminal variance increase occurs at $\epsilon=0.5$ when $\omega=0.001$ and 0.003, and at $\epsilon=0.1$ when $\omega=0.01$. However, at $\epsilon=1$, the improvements are smaller due to the network's increased resistance to polarization under stronger attractive opinion dynamics. Similar trends are observed in the Gilets Jaunes dataset. For minimizing variance, $\epsilon=0.1$ leads to the smallest terminal variance, while for maximizing variance, $\epsilon=0.3$ results in the largest terminal variance. 

These findings provide useful guidance when designing shadow banning policies.  For objectives involving the opinion mean, the precise choice of opinion dynamics parameters is not critical.  For objectives involving the opinion variance, one must decide if the opinion dynamics shows strong or weak persuasion, as strong persuasion makes shadow banning harder to overcome the natural attractive opinion dynamics. Since real-world social networks exhibit persistent polarization, better shadow banning policies will be calculated if using opinion dynamics models with weak persuasion.

\clearpage
% \bibliography{YS}

\end{document}